\documentclass[a4paper,12pt]{scrartcl}
\pdfoutput=1
\usepackage{lmodern}
\usepackage{amssymb,amsmath}
\usepackage{fixltx2e} 
\usepackage[utf8]{inputenc}
\makeatletter\AtBeginDocument{%
  \renewcommand{\@listi}
    {\setlength{\labelwidth}{4em}}
}\makeatother
\usepackage{enumerate} 
\usepackage{ctable}
\usepackage{float} 
\usepackage{graphicx}
\usepackage{url}
\usepackage{comment}
\usepackage[american]{babel}
\usepackage{xspace}
\usepackage{subcaption}
\usepackage{paralist}
\usepackage{multirow}
\usepackage{ifthen}
\usepackage{setspace}
\usepackage[np]{numprint}
\usepackage{pgf}
\usepackage{fancybox}
\usepackage{alphalph} 

\makeatletter
\let\@fnsymbol\@alph
\makeatother

\begin{document}

\definecolor{ffi}{rgb}{0.698,0.122,0.435} 
\definecolor{rja}{rgb}{0.122,0.435,0.698} 
\definecolor{chu}{rgb}{0.435,0.698,0.122} 
\definecolor{TODO}{rgb}{0.784,0.145,0.00}
\newcommand{\todo}[2][TODO]{\textcolor{#1}{\textbullet}\marginpar{\color{#1}\Ovalbox{\parbox{.90\marginparwidth}{\footnotesize\raggedright \textbf{#1:} #2}}}}
\newcommand{\komm}[2]{\textsf{\bfseries\color{#1}/* #2 (#1) */}}

\newcommand{\rja}[1]{\komm{rja}{#1}}\newcommand{\mrja}[1]{\todo[rja]{#1}}
\newcommand{\ffi}[1]{\komm{ffi}{#1}}\newcommand{\mffi}[1]{\todo[ffi]{#1}}
\newcommand{\chu}[1]{\komm{chu}{#1}}\newcommand{\mchu}[1]{\todo[chu]{#1}}
\newcommand{\final}[1]{\textbf{/* for camera ready/long version: #1  */}}
\newcommand{\journalKomm}[1]{\textbf{/* for journal version: #1  */}}
  \renewcommand{\komm}[2]{}\renewcommand{\todo}[2][]{}
  \renewcommand{\journalKomm}[1]{}
  \renewcommand{\final}[1]{}

\definecolor{newColor}{rgb}{0.00,0.145,0.784}
\newcommand{\neu}[1]{\textcolor{newColor}{#1}}

\newcommand{\plosincludegraphics}[2][]{\includegraphics[#1]{#2}}

\ifthenelse{\equal{\jobname}{\detokenize{plosfinal}}}{
  \renewcommand{\komm}[2]{}\renewcommand{\todo}[2][]{}
  \renewcommand{\journalKomm}[1]{}
  \renewcommand{\plosincludegraphics}[2][]{}
}{
  \renewcommand{\final}[1]{}
}

\hyphenation{Page-Rank Eng-lish ma-nipu-lated know-ledge DB-pedia}

\newcommand{\ie}{i.e.,\xspace}
\newcommand{\eg}{e.g.,\xspace}
\newcommand{\furl}[1]{\footnote{\url{#1}}}
\newcommand{\class}[1]{\textsf{#1}\xspace}

\setcounter{topnumber}{5}
\setcounter{bottomnumber}{5}
\setcounter{totalnumber}{20}
\renewcommand{\topfraction}{1}
\renewcommand{\bottomfraction}{1}
\renewcommand{\textfraction}{0}
\renewcommand{\floatpagefraction}{0.99}
\renewcommand{\dbltopfraction}{1}
\renewcommand{\dblfloatpagefraction}{0.99}


\newcolumntype{L}[1]{>{\raggedright\let\newline\\\arraybackslash\hspace{0pt}}p{#1}}
\newcolumntype{C}[1]{>{\centering\let\newline\\\arraybackslash\hspace{0pt}}p{#1}}
\newcolumntype{R}[1]{>{\raggedleft\let\newline\\\arraybackslash\hspace{0pt}}p{#1}}

\newcommand{\wpfile}[1]{{\sffamily \MakeLowercase{#1}}} 
\newcommand{\wptype}[1]{\emph{#1}} 
\newcommand{\wpprop}[1]{\emph{#1}} 
\newcommand{\wpcat}[1]{\emph{#1}} 
\newcommand{\wpent}[1]{\emph{#1}} 
\newcommand{\apInfoboxesEn}{template \emph{Writer}\xspace} 
\newcommand{\apCategoryGraph}{category \emph{Writers}\xspace} 
\newcommand{\apInfProps}{property \emph{Occupation}\xspace} 

\newcommand{\rpagelen}{{\scshape PL}\xspace}
\newcommand{\rinlinks}{{\scshape IL}\xspace}
\newcommand{\rprwriter}{{\scshape PW}\xspace}
\newcommand{\rprcomplete}{{\scshape PC}\xspace}
\newcommand{\rpvtwo}{{\scshape V12}\xspace}
\newcommand{\rpvthree}{{\scshape V13}\xspace}
\newcommand{\rpvfour}{{\scshape V14}\xspace}
\newcommand{\x}{$\times$}

\title{World Literature According to Wikipedia: Introduction to a DBpedia-Based Framework}
\author{
Christoph Hube,\textsuperscript{1}
Frank Fischer,\textsuperscript{2}
Robert Jäschke,\textsuperscript{1,3}\\
Gerhard Lauer,\textsuperscript{4}
Mads Rosendahl Thomsen\textsuperscript{5}}
\date{\small
\textsuperscript{1} L3S Research Center, Hannover, Germany\\
\textsuperscript{2} National Research University Higher School of Economics, Moscow, Russia\\
\textsuperscript{3} University of Sheffield, United Kingdom\\
\textsuperscript{4} Göttingen Centre for Digital Humanities, Germany\\
\textsuperscript{5} Aarhus University, Denmark
}
\maketitle

\begin{abstract}
  Among the manifold takes on world literature, it is our goal to
  contribute to the discussion from a digital point of view by
  analyzing the representation of world literature in Wikipedia with
  its millions of articles in hundreds of languages.  As a
  preliminary, we introduce and compare three different approaches to
  identify writers on Wikipedia using data from DBpedia, a community
  project with the goal of extracting and providing structured
  information from Wikipedia. Equipped with our basic set of writers, we
  analyze how they are represented throughout the 15 biggest Wikipedia
  language versions. We combine intrinsic measures (mostly examining
  the connectedness of articles) with extrinsic ones (analyzing how
  often articles are frequented by readers) and develop methods to
  evaluate our results. The better part of our findings seems to
  convey a rather conservative, old-fashioned version of world
  literature, but a version derived from reproducible facts revealing
  an implicit literary canon based on the editing and reading behavior
  of millions of people. While still having to solve some known
  issues, the introduced methods will help us build an observatory of
  world literature to further investigate its representativeness and
  biases.
\end{abstract}
\section{Introduction}\label{sec:introduction}


Ever since Johann Wolfgang von Goethe introduced his conception of
`world literature' in one of his conversations with Eckermann, in
January, 1827 \cite{birus2004goethe}, the term refers to authors and
works that transcend national and language borders. While this might
still be the least common denominator, opinions on what belongs to world
literature have become very diverse in recent years, bringing several
factors into play. In their introduction
\cite{levine2013what} to a 2013 special issue of \emph{Modern Language
Quarterly} entirely dedicated to the question ``What Counts as World
Literature?'', Levine and Mani ask: ``Is world literature simply a
prerogative of the professional reader, the literary theorist, or is
it a much larger interactive space with numerous actors who range from
authors, translators, and readers to librarians, publishers,
collectors, and booksellers?''

The aim of this paper is to introduce another way to explore what
counts as world literature in a variety of contexts.  If ``world
literature as a publishing and teaching project was part of a push to
democratize high culture in the early twentieth
century'' 
\cite{levine2013what}, then the launch of Wikipedia in 2001 can be
understood as a project that democratized the gathering and
distribution of general knowledge by harnessing the much-written-about
``wisdom of the crowd''. The number of articles in the English
Wikipedia version
is, to date, two magnitudes higher than that of the last print edition
of the Encyclopædia Britannica, and our idea is to show how world
literature is represented in this vast digital resource. To pursue
this goal, we make use of DBpedia, one of several attempts to
formalize the contents of Wikipedia by converting a human-readable,
hypertextual encyclopedia into a machine-readable, queryable
database.



The individual steps described in this paper include:
\begin{itemize} 
\item A comparison of three different approaches for identifying
  writers of literature across different language versions of Wikipedia.
\item An evaluation of our set of extracted writers, including
 their temporal distribution.
\item A comparison and evaluation of five different intrinsic and
  extrinsic ranking measures to assess the importance of writers on
  Wikipedia.
\item A visualization of the network of the most important writers, a
  minimum definition of what world literature is, according to
  Wikipedia.
\item An approach to identify writers who transcended language
  boundaries, in the 15 most comprehensive Wikipedia language
  editions, including a detailed analysis.
\item Publication of additional results (tables and datasets) on
  \url{http://data.weltliteratur.net/}, our project page that is planned
  to become an observatory of the digital discourse on world
  literature.
\end{itemize}

The paper is structured as follows: In Section~\ref{related-work} we
discuss possible definitions of world literature and briefly introduce
Wikipedia and DBpedia.
In Section~\ref{sec:datasets} we explicate the used DBpedia datasets
and the criteria for the selection of the 15 Wikipedia language
editions. We introduce our approaches for the identification of
writers in Section~\ref{sec:approaches}.
In Section~\ref{sec:analysis} we explain the dataset creation and
subsequently present our results in
Sections~\ref{sec:most-import-writ} and
\ref{sec:native-writers}. Section~\ref{sec:conclusion} wraps up
this paper drawing conclusions from our results and trying to
cautiously describe Wikipedia's inherent idea of world literature.

\section{Related Work}\label{related-work}

In this section, we briefly introduce our understanding of the ongoing
discourse on world literature. We also present the required technical
background and introduce projects that were important to conducting
our own research.

\subsection{What is World Literature?}

Definitions of world literature are manifold,
and while postmodernism, postcolonial and gender studies
significantly diversified our perception of world literature,
there is still one consensual aspect that can be filtered
out of ongoing discussions and according to which world
literature comprises ``all literary works that
circulate beyond their culture of origin, either in translation
or in their original language'' \cite{damrosch2003what}.
This aspect does not seem to have changed since the term was
coined by Goethe. However, when it comes to measuring the
global significance of a work or author, opinions differ.
Is it the number of translations of a work that counts (something
that could be operationalized using the Index Translationum database)?
Are sales important or is it an indication of questionable literary
qualities if a literary work hits best-seller lists? It is difficult
to find common grounds as world literature is an ever-changing entity
depending on a continued conversation on value and influence.

While the formation of a world literary canon is an underlying
part of literary criticism whenever a work or an author is selected
instead of others, most notably and idiosyncratically in the works
of Harold Bloom, world literature can also be studied based on the
degree of international reception of certain works and authors.
If we follow David Damrosch who makes his case for world literature as
``a mode of circulation and of reading'' \cite{damrosch2003what},
there still remains the problem of measurement.
With our author-centric
approach, we determine the degree of reception of a writer by measuring
his or her presence in the different language editions of Wikipedia,
a new spin on the question if and how an author crossed national
and language borders.






\subsection{Research on Wikipedia}\label{sec:wikipedia}

Since its launch in 2001, the free-access and free-content internet
encyclopedia Wikipedia has become the web's largest and most popular
general reference work, ranked among the top ten most popular
websites. More than 60,000 active editors work on articles in editions
for more than 200 different languages. Among these editions, the
English Wikipedia sticks out with its more than 4.7 million
articles at the beginning of 2015.

By now, Wikipedia is widely approved as a resource for scientific
research.  The encyclopedia has been called a ``global memory place''
\cite{pentzold2009floatingGap} and a ``goldmine of information''
\cite{Medelyan2009Mining}, emphasizing its value for researchers.
Nielsen \cite{Nielsen2013WikipediaResearchReport} presents various
research results and groups them into several main categories,
differentiating between research that examines Wikipedia and research
that uses information from Wikipedia to draw conclusions about other
matters.
%
%
Also in the Digital Humanities, Wikipedia is regarded as an important
research subject, be it by analyzing intellectual connections among
philosophers \cite{athenikos2009philosophers} or evaluating the
significance of historical entities
\cite{takahashi2011historicalEntities} based on infobox properties
and page links.
Another example is the application of social network analysis to
famous persons on Wikipedia 
\cite{Aragon2012biographical} by ranking them using the in-degree,
out-degree, and PageRank of their articles, to show differences and
similarities between different Wikipedia language versions. 
Using three different ranking algorithms, Eom et al.
\cite{eom2015interactions} determine the top historical 
figures of 24 Wikipedia language versions and their evolution over time.
Gloor et al. \cite{gloor2015cultural} use a similar approach ranking
historical persons according to their influence on other persons 
during their own lifetime.
More recently, Laufer et al. \cite{laufer2015mining} analyzed how
European food cultures are represented in different Wikipedia
editions.
%
%
However, similar analyses for writers or literature have not been
performed so far and the existing works do not comprehensively
evaluate different approaches for identifying and ranking persons.

\subsection{Quality Aspects of Wikipedia}
\enlargethispage{\baselineskip}

While some early studies (\eg \cite{giles2005nature}) already
suggested that the quality of Wikipedia is comparable to that of
commercial encyclopedias, there has also been severe criticism
(\eg \cite{black20110wikipedia}). Yet, in its 14 years of
existence the Wikipedia project has been constantly evolving and
has seen the introduction of several quality assurance measures.
%
%
A good overview on papers, articles, and studies on different
dimensions of the quality of Wikipedia is given by Nielsen
\cite{Nielsen2013WikipediaResearchReport}. The dimensions comprise,
among others, accuracy, coverage, bias, conciseness, and
up-to-dateness, and especially its up-to-dateness and multilingualism
are regarded as major strengths
of Wikipedia \cite{hammwoehner2007Qualitaetsaspekte}.
A research topic on its own is the diversity between Wikipedia
language versions
\cite{Bao2012Omnipedia,hecht2010babel,Callahan2011culturalbias} with
findings showing that Wikipedia has a cultural bias also when it comes
to portraying famous persons \cite{Callahan2011culturalbias}.
Our results also suggest that the popularity of a writer within a
Wikipedia language version depends on whether the writer is associated
with the particular language. Wikipedia can therefore be seen as a
``socially produced document'' that represents the values and
interests of the people who use it \cite{royal2009Completeness}. As we
will see in Section~\ref{sec:writers-their-own} this can as well lead
to situations in which single persons gain unusually high attention in
some language versions due to the efforts of single editors.

Halavais and Lackaff \cite{halavais2008TopicalCoverage} analyze the
topical coverage of Wikipedia by examining a randomly-drawn set of
3,000 articles. They discover that Wikipedia is, like many other
encyclopedias, not as strong in the 
humanities as it is in the natural
sciences, though there actually exists a large number of articles
representing literary criticism, especially regarding fiction.
A quality study of Wikipedia examining the representation of the works
of William Shakespeare in the English and German Wikipedia editions
found that the articles in the English edition are of higher quality
than their counterparts in the German edition
\cite{hammwoehner2007Qualitaetsaspekte}.

Although these and 
other papers have analyzed different aspects of the quality of the
representation of literature on Wikipedia as part of
general 
quality analyses, there are, to the best of our knowledge, no works
specifically targetting writers or literature. Furthermore,
our approach does not focus on the quality aspect, but in fact
considers Wikipedia data as a representation of world literature
from the point of view of expert or non-expert editors and readers.


\subsection{DBpedia}\label{sec:dbpedia}

\begin{figure}
  \centering
  \plosincludegraphics[height=124mm]{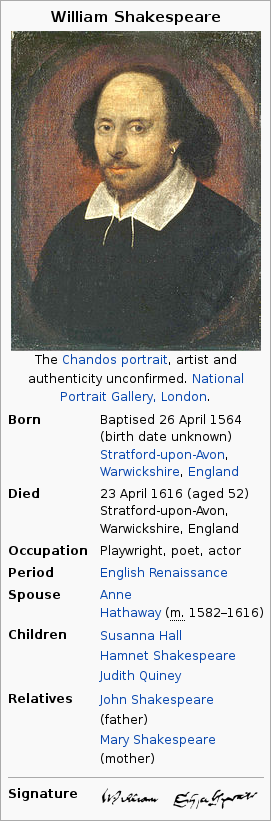} \hspace{2em} 
  \plosincludegraphics[height=124mm]{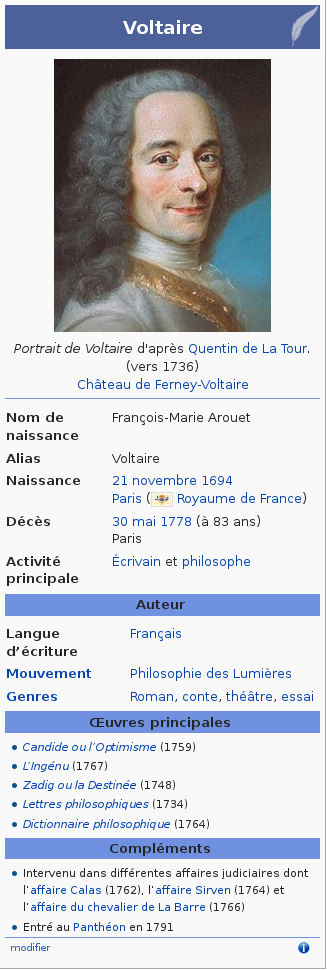}
  \caption{Infoboxes for \wpent{William Shakespeare} from the English Wikipedia and \wpent{Voltaire} from the French Wikipedia.}
  \label{fig:shakespeare}
\end{figure}

The crowdsourced community project DBpedia
\cite{Auer2007dbpediaNucleus} provides a knowledge base of rich data
extracted from Wikipedia.
The entire DBpedia knowledge base describes facts for more than 38
million ``things'' from 125 Wikipedia language versions. Part of the
project is the crowdsourced construction of an ontology by manually
extracting infobox properties of different language versions (a
Wikipedia infobox is a table in the top-right corner of an article
presenting a subset of structured information in the form of
attribute--value pairs; \eg infoboxes on persons will usually contain
data fields like ``Born'', ``Died'', or ``Occupation'', see
Fig.~\ref{fig:shakespeare}).
Since its initial release in 2007, DBpedia has been used many times as
a foundation for research with and about Wikipedia
\cite{nitzschke2012academicImpact}.
%
%
One of the major advantages of DBpedia is the possibility to query
data from the infoboxes of articles, which is not possible from within
Wikipedia. This enables us to extract specific writer sets, \eg `all
writers who were born after 1910'.
%
However, due to different options for labeling articles about writers
across different Wikipedia language versions, the
identification of writers in the DBpedia datasets is not
trivial. We test three approaches to identify writers in
Section~\ref{sec:approaches}.
In an effort to simplify the handling of data across language versions
and to enable rich queries similar to DBpedia, the Wikimedia
foundation has started the Wikidata project
\cite{vrandecic2012wikidata} which provides a database for infobox
(and other) data. However, at present, infoboxes are still curated
within Wikipedia and therefore Wikidata can not yet substitute
DBpedia.


\section{DBpedia Datasets and Selection of Language Versions}\label{sec:datasets}

The DBpedia 2014 download is based on the Wikipedia dump from late
April/early May 2014 \cite{dbpedia2014download}. 
The files belonging to the datasets from a
specific language version can be found in folders
named with their respective language codes. All
datasets are available in several RDF serializations (\eg N-Triples
and Turtle) and are compressed with bzip2.
For example, the \wpfile{page\_links} dataset of the French Wikipedia
can be found at the URL
\url{http://downloads.dbpedia.org/2014/fr/page_links_fr.nt.bz2}.
The following list shows all DBpedia datasets we use in this work:
\begin{description}\setlength{\itemsep}{0pt}
\item[\wpfile{article\_categories}:] maps articles to their
  categories.
\item[\wpfile{infobox\_properties}:] properties extracted
  from the infoboxes of articles.
\item[\wpfile{instance\_types}:] maps articles to their types
  according to templates used within the article, for example
  a \wptype{Writer} infobox.
\item[\wpfile{interlanguage\_links}:] maps articles to their
  counterparts in other language versions.
\item[\wpfile{mappingbased\_properties}:] a cleaned version of the
  \wpfile{infobox\_properties} dataset.  For example, the properties
  \wpprop{birthYear} and \wpprop{yearOfBirth} are both mapped to the
  property \wpprop{birthYear}.
\item[\wpfile{page\_length}:] the number of characters for each
  article.
\item[\wpfile{page\_links}:] links between articles in the same
  language version.
\item[\wpfile{skos\_categories}:] the Wikipedia category graph
  containing the links between categories.
\end{description}

Since DBpedia follows a crowdsourcing approach based on extraction
rules written by community members, not all datasets are
available for all language versions. Some datasets, like
\wpfile{instance\_types}, rely on the DBpedia ontology data. According
to \cite{lehmann2014dbpediaKnowledgeBase}, the DBpedia ontology
currently comprises mappings for infobox data for 27 language
versions.

%
%
Due to this restriction and also for practical purposes, we have to
restrict our analysis to a subset of the 125 languages available in
DBpedia and look at two criteria: the size of the version, \ie
the \emph{number of articles}, and the Wikipedia \emph{article
  depth} \cite{wikipediarticledepth}. 
The article depth is a measure of quality for a language version. A
high depth indicates that articles are frequently updated, meaning
that the average number of edits per article is relatively high.
%

In addition to articles created by humans, many Wikipedia versions
also contain articles that were automatically generated by
\emph{bots}, \eg by copying and translating (parts of) existing
articles from other languages. For instance, the Waray language spoken
in the Philippines has less than 3 million native speakers, but the
Waray Wikipedia
\cite{wp:waray} 
is among the top ten editions when just counting the number of
articles. The majority of these articles were generated by a bot, the
average number of edits per article is less than 3 (for comparison,
the Finnish edition ranks 20th and has around 42 edits per article on
average).
The use of bots has been criticized by the community \cite{bots:digitopoly,steiner2014bots} 
%
and also for our purpose of analyzing the perception of world
literature, a too-large presence of bot-generated articles
would be counterproductive.
By taking the article depth into account we make sure that language
versions with a high percentage of bot-generated articles are not too
dominant.

We decided to consider the 15 language editions that were both among
the top 30 by number of articles and by article depth as at
March~7, 2015 (cf. Table~\ref{tab:languages}).
Although the ranking of language editions could
  have changed since the creation of the DBpedia dump, this would only
  affect our choice of languages but not the general results of our
  analysis.
%
The \#writers column depicts the number of articles of the type
\wptype{Writer} according to the \wpfile{instance\_types} dataset.
Language versions for which no such dataset exists are indicated by
`--'.  Even though there exist \wpfile{instance\_types} datasets for
the German and Italian Wikipedia, these editions lack an equivalent to
the English \wptype{Writer} type. Therefore, the number of writers for
these two language versions equals zero.
The last column (\#writers in our set) is explained in
Section~\ref{sec:choosing_approach}.
%
%

\nprounddigits{0}

\begin{table}
  \caption{\bf Properties of the 15 Wikipedia language versions selected.}
  \begin{tabular}{@{}llrrrrr@{}}
    \toprule
    language       &    code &    \#articles  &    depth  &    \#edits       &    \#writers    &  \#writers in our set \\ 
    \midrule
    English        &    en     &    4,738,257 &    \np{895.18} &    759,894,686 &    25,995       &  10,765 \\ 
    German         &    de     &    1,822,322 &    \np{ 88.32} &    145,142,072 &    0            &	5,202  \\ 
    French         &    fr     &    1,598,081 &    \np{196.93} &    114,926,655 &    7,974        &	4,981  \\ 
    Russian        &    ru     &    1,197,015 &    \np{116.73} &    81,444,167  &    11,515       &	3,286  \\ 
    Italian        &    it     &    1,177,497 &    \np{107.40} &    76,520,585  &    0            &	3,514  \\ 
    Spanish        &    es     &    1,161,825 &    \np{177.01} &    85,947,882  &    6,668        &	3,430  \\ 
    Portuguese     &    pt     &    867,553   &    \np{124.47} &    42,572,971  &    1,104        &	2,226  \\ 
    Chinese        &    zh     &    814,358   &    \np{134.23} &    35,959,378  &    --           &	1,468  \\ 
    Persian        &    fa     &    446,641   &    \np{209.27} &    19,684,602  &    --           &	1,395  \\ 
    Arabic         &    ar     &    352,720   &    \np{202.99} &    17,792,878  &    1,288        &	1,017  \\ 
    Korean         &    ko     &    306,641   &    \np{ 85.53} &    15,571,911  &    968          &	1,052  \\ 
    Hungarian      &    hu     &    286,466   &    \np{ 91.84} &    16,382,712  &    3,033        &	1,149  \\ 
    Serbo-Croatian &    sh     &    278,382   &    \np{354.65} &    9,143,222   &    --           &	849    \\ 
    Romanian       &    ro     &    271,118   &    \np{ 93.13} &    9,489,089   &    --           &	1,617  \\ 
    Turkish        &    tr     &    242,778   &    \np{205.94} &    16,123,456  &    1,314        &	1,217  \\ 
    \bottomrule
  \end{tabular}
  \label{tab:languages}
\end{table}

\section{ Identifying Writers on Wikipedia}\label{sec:approaches}

The automatic extraction of writers in DBpedia is a non-trivial task
as there is neither a comprehensive nor coherent classification of
writers across language versions. 
We experimented with several ideas to address this challenge, each of
which leads to a different set of writers. As we will see, the
decision for an approach depends on the question of who is considered
a writer. We tested the following three approaches for writer
identification:
\begin{description}\setlength{\itemsep}{0pt}
\item[\apInfoboxesEn:] articles featuring the \wptype{Writer} template
  %
\item[\apCategoryGraph:] articles belonging to the \emph{Writers} category in
  the category graph
\item[\apInfProps:] articles featuring a writer-related \wpprop{Occupation}
  infobox property
\end{description}
We first describe the three approaches and their application to the
English Wikipedia and then discuss whether and how they can be applied
to other language versions. We also evaluate and discuss which
approach meets our requirements better than the others.


\subsection{\emph{Writer} Template}\label{sec:writer_infoboxes_english}

DBpedia provides \wpfile{instance\_types} datasets for several
language editions which map instances to types. Each instance
represents an article in the corresponding edition. An instance is
classified as \wptype{Writer}, if the corresponding article contains a
specific kind of template which in most cases is an infobox of the
type \wptype{Writer}.
%
%
As an example, Fig.~\ref{fig:shakespeare} shows the infobox for
\wpent{William Shakespeare} from the English
Wikipedia 
and the infobox for \wpent{Voltaire} from the Wikipédia en français.
%
%
Both infoboxes contain basic information like name, day of birth,
day of death, and occupation. The French infobox further
contains writer-specific information like writing
languages, repertoire of genres, and important works.
In addition, the French \wptype{Writer} infobox is marked with a
pinfeather in the upper right corner to indicate that a person
is a writer.
All properties in the infoboxes are optional and many of them are
general person properties, like \wpprop{birthDate} and
\wpprop{deathDate}. Exploiting the \wpfile{instance\_types\_en}
dataset we are able to extract all articles in the English Wikipedia
containing a \wptype{Writer} template.

After building this \emph{basic set} of persons with \wptype{Writer}
templates in the English Wikipedia, we use this set to identify
writers in other language versions. By checking which of the persons
in the English set are also featured in other language versions, we
obtain subsets for all language versions. The articles in other
language versions do not have to contain a \wptype{Writer}
template. Thereby we bypass the problem that \wpfile{instance\_types}
datasets are only available for a small number of language
versions. The drawback of this approach is, however, that we omit
writers that are neither mentioned nor classified as such in the
English Wikipedia.


We also considered each language version separately. This would
allow us to find writers without a \wptype{Writer} template or even
without an article in the English Wikipedia, as long as their article
in the particular language version contains some kind of
\wptype{Writer} template.
%
Unfortunately, DBpedia does not provide \wpfile{instance\_types}
datasets for many language versions and some language versions do
not provide \wptype{Writer} templates. For example, in the German
Wikipedia, writers are described through a simple (hidden)
\wptype{Person} infobox and thus can not be classified as writers in the
\wpfile{instance\_types\_de} dataset. As a consequence, these writers
are not recognized and the number of writers found for some language
versions equals zero (see Table~\ref{tab:languages}). The DBpedia
community already dealt with this problem by providing specific
datasets containing data extracted from the running text of the
article to classify persons. Although we appreciate this effort, we
decided not to use this data for our research to avoid a bias by
mixing different extraction methods.

\subsection{Traversal of the Category Graph}\label{sec:category_graph}

Wikipedia provides a facility to assign articles to
\emph{categories}. Accordingly, the article on \wpent{William
  Shakespeare} belongs to 19~categories including the categories
\wpcat{16th-century English
  Writers} 
and \wpcat{17th-century English
  Writers}. 
Categories can be assigned to other categories, forming the Wikipedia
\emph{category graph}. In this graph topically similar categories are
grouped together. Both \wpcat{16th-century English Writers}
and \wpcat{17th-century English Writers} are a
subcategory of \wpcat{English Writers by Century}. 
Using the \wpfile{article\_categories} dataset from DBpedia we are
able to extract all articles belonging to a particular category while
the \wpfile{skos\_categories} dataset contains the category graph
itself. The challenge we face is to find a way through the graph that
will essentially give us articles on writers without delivering too
many non-writer articles.
Such an approach highly depends on the choice of the root category as
starting point for the traversal and appropriate termination and filtering
conditions.

Unfortunately, the assignment of articles to categories is
inconsistent. To obtain the majority of writers we need to cover a
large number of categories within a breadth-first search approach
resulting in a larger number of articles that are not primarily about
writers. 
Starting the traversal at the \wpcat{Writers by Century} category in
the English version, our set contained persons like \wpent{Winston
  Churchill} and \wpent{Leonardo da Vinci}, 
while omitting writers like \wpent{Gertrude Stein} and \wpent{Heinrich
  Heine} (Table~\ref{tab:topfive_combined}).
The selection of the \wpcat{Writers} category as root category even
increased the amount of non-writers, so we started to add filters to
omit categories that were not directly related to writers, \eg by
involving only categories that contain the word `writer'
(writer filter).
We also realized that we needed to add some kind of termination
condition, since many famous writers were represented not by an
article alone but by a category containing their works, etc.  In
addition, the category graph can contain cycles, so we had to make
sure that each category was consulted only once. To ensure that the
resulting set only contains persons, we compared all instances to
DBpedia's persondata set.

Even by trying several different root categories and filters we were
not able to extract an appropriate set of writers from the category
information.
The difficulties that occur while using the Wikipedia category graph
have been described before, \eg in \cite{Medelyan2009Mining}.

The benefit of this approach is that it can be applied to
non-English language versions. However, the category graphs
tend to be different in different editions, adjusting them would
take individual effort. One would need to identify the appropriate
root category of each language version and verify the
consistency of its subcategories. Even if the name of a category
corresponds to a category in the English graph, it does not mean
that the category is used in the same manner. If we would go with
this assumption anyhow, it could deliver strongly inconsistent sets
from the different language versions.

\subsection{\emph{Occupation} Infobox Property}\label{sec:infobox_properties}

As we have seen in Section \ref{sec:writer_infoboxes_english},
the type of an infobox can be used to classify an article. Another
approach to identify writers on Wikipedia is to use the property
values that are contained in the infobox of an article. For instance,
the infobox of \emph{William Shakespeare} in the English Wikipedia,
depicted in Fig.~\ref{fig:shakespeare}, contains an \emph{Occupation}
property. In the case of \emph{Shakespeare}, the value of this
property is ``Playwright, poet, actor''. The infobox of
\emph{Voltaire} in the French Wikipedia contains a similar property
called \emph{Activité principale} (main activity).

In the English Wikipedia the \emph{Occupation} property is used quite
frequently: The \wpfile{infobox\_properties} dataset contains 240,994
instances using the property, 32,722 of them include the term
``writer'' in the value of this property. For the extraction of
writers it is useful to add further terms to the query, like ``poet''
and ``novelist''. We also excluded particular terms like
``songwriter'' and ``screenwriter'' (as stressed before, these
are contingent decisions, but we had to draw a line somewhere
to operationalize our research).

This approach can be used for every Wikipedia language version that
uses a similar infobox property, but the search terms have to be
defined for every version separately. This might be a problem since we
do not know whether the translations of terms are used in the same way
in the respective language version as the English terms are used in
the English Wikipedia. Of course, the idea of building a basic set of
writers in the English Wikipedia and taking it from there can also be
used for this approach.

\subsection{Comparison of the Approaches}\label{sec:choosing_approach}

Our opting for one of the introduced approaches is influenced by
several questions:
\begin{enumerate}[1.] 
\item What is a writer? Should the set contain persons who are not
  primarily considered writers? Do we need to
  distinguish between different types of writers (e.g., fiction,
  non-fiction)? 
\item Do we want to focus on precision or recall, \ie should the
  set contain as few non-writers as possible or as many writers as
  possible?
\item Which language versions should be considered?
\item Which approach could be suitable to extract literary works?
\end{enumerate}


%
%
\newcommand{\persct}[1]{\multicolumn{2}{c}{(#1 articles)}}
\begin{table}
  \caption{\bf Comparison of writer identification approaches.}
  \label{tab:topfive_combined}
  \begin{tabular}{@{}p{19mm}lrlr@{}}
    \toprule
    & \multicolumn{2}{c}{a) only by this approach} & \multicolumn{2}{c}{b) missing by this approach} \\
    \cmidrule(rl){2-3}\cmidrule(l){4-5}
    approach & \multicolumn{2}{c}{person\hfill\#in-links} & \multicolumn{2}{c}{person\hfill\#in-links}\\
    \midrule
    template          &Robert Christgau        &5,844 &Arnold Schwarzenegger &2,255	\\
    \emph{Writer}     &James Berardinelli      &409   &Tupac Shakur          &1,591	\\
                      &Carlo Goldoni           &380   &Geoffrey Chaucer      &1,250	\\
                      &Constantin Stanislavski &338   &Franz Kafka           &1,229	\\
    \#p:\hfill 25,995 &John Osborne            &291   &William Shatner       &1,184	\\
    \#N:\hfill 102    &\persct{5,504}          & \persct{633}\\ 
    \midrule
    category          &Bob Dylan         &5,996 &P.\,G. Wodehouse           &941	\\
    \emph{Writers}    &George Washington &5,305 &Robert Graves              &662	\\
                      &Paul McCartney    &4,619 &Bram Stoker                &647	\\
                      &Elton John        &4,553 &Oliver Goldsmith           &424	\\
    \#p:\hfill 95,541 &John Lennon       &4,487 &Algernon Charles Swinburne &409	\\
    \#N:\hfill 94     &\persct{62,310}   & \persct{300}\\ 
    \midrule
    category          &Winston Churchill &5,208 &Rabindranath Tagore   &1,324\\
    \emph{Writers by} &Gautama Buddha    &3,114 &Alfred, Lord Tennyson &1,145\\
    \emph{Century}    &Leonardo da Vinci &1,975 &Walter Raleigh        &752	\\
                      &Strabo            &1,620 &Gertrude Stein        &608	\\
    \#p:\hfill 28,245 &Maimonides        &1,129 &Heinrich Heine        &582		\\
    \#N:\hfill 57     &\persct{10,898}   & \persct{5,272}\\ 
    \midrule
    property          &Tom Cruise     &1,416 &Roger Ebert         &3,924	\\
    \emph{Occupation} &John Travolta  &1,128 &Edgar Allan Poe     &2,206	\\
                      &Tim Burton     &1,032 &Stephen King        &2,199	\\
                      &Bruce Willis   &960   &George Bernard Shaw &1,812	\\
    \#p:\hfill 18,534 &Buster Keaton  &795   &Ernest Hemingway    &1,764	\\
    \#N:\hfill 76     &\persct{3,589} & \persct{2,061}\\ 
    \bottomrule
  \end{tabular}
  \begin{flushleft}
    For each approach we list the top five writers (according to the
    number of incoming links to their articles) that are a) only
    identified by this approach, or b) missing by this approach but
    identified by all other approaches. The first column shows for
    each approach the number of persons (\#p) and the number of Nobel
    laureates in literature (\#N) it identifies.
  \end{flushleft}
\end{table}


To obtain an impression of the different approaches,
Table~\ref{tab:topfive_combined} shows a comparison.
For the category-based approach we show the results using two
different root categories, namely \wpcat{Writers by Century}\xspace
and \wpcat{Writers}.
%
%
For each approach the table lists the top five writers \emph{only}
identified by the corresponding approach and those \emph{not}
identified by the same approach (including the total number of
articles in each of the two sets). In addition, the first column lists
the overall number of Wikipedia articles that were identified as
writers by each approach and the number of Nobel laureates in
literature contained in the respective set (we know that using the
list of Nobel laureates as some kind of fallback canon is an audacious
venture and that this list is far from being a gold standard, also
because it was only started in the 20th century, but
it proved insightful when evaluating our approaches). 
(For a more
  detailed description and analysis of the Nobel-laureate data
  cf. Section~\ref{sec:nobel-laureates}.) 
%

%
%
While the top five authors show which kind of persons are
considered writers by each approach (or not), the numerical values
in the first column allow us to quantitatively assess the
performance. 
%
%
%
%
%
%
Comparing the number of identified persons (\#p) against the number of
Nobel laureates in literature (\#N), we can see that the
\apInfoboxesEn approach yields the largest number of Nobel
laureates (102 of 111) within a modest number of persons (25,995).
Plus, only few articles (633) could not be found by the
\apInfoboxesEn approach but by the other approaches. Although the
\apCategoryGraph approach misses even less articles (300), it does so at
the cost of a much higher number of detected persons (95,541) and a
still smaller number of correctly identified Nobel laureates.
The other two approaches are much worse in correctly identifying
Nobel laureates and also have a larger number of persons that they do
not identify as writers.
Altogether, the \apInfoboxesEn approach is clearly the most selective
approach. 
One reason for the good performance of this method could be the fact 
that no \wptype{Writer} article in our dataset contains more than
one template and is therefore not assigned to more than one type.
In fact, it is very unusual for a Wikipedia article to contain more
than one template. Thus, only articles about persons who are primarily
considered writers by Wikipedia editors are equipped with the
\wptype{Writer} template. Contrariwise, an article can have several
categories and the \wpprop{Occupation} property can contain several
occupations.


From a qualitative point of view, the top five persons in
Table~\ref{tab:topfive_combined} provide a mixed
impression. Apparently, singer-songwriters are frequently categorized
as \wpcat{Writers} (which is not untrue, of course, but lacks
the distinctiveness we need for our operationalization).
In the \apInfProps approach we could exclude them by blacklisting the
term ``songwriter''. On the other hand, this approach identifies
a bunch of popular actors as writers who apparently have not been categorized
as \emph{Writers}.
Reasons may vary, some might have written an autobiography or another
kind of book or coauthored a screenplay.
The \apInfoboxesEn approach is the only one identifying critics
\emph{Robert Christgau} and \emph{James Berardinelli} as
writers (which, again, is not untrue, but not helpful either).
The top persons not identified by the corresponding approaches
clearly show that there is room for improvement: every approach
misses important writers, e.g., \emph{Franz Kafka} has no
\apInfoboxesEn, \emph{P.\,G. Wodehouse} no \apCategoryGraph,
Nobel prize laureate \emph{Rabindranath Tagore} lacks the
\emph{Writers by Century} category, and \emph{Edgar Allan Poe} a \apInfProps.

\subsection{Discussion and Selection of an Approach}\label{sec:disc-select-an}

We decided to aim at a high precision and include only persons who are
primarily considered writers, which correlates with the decision to
prefer writers of fiction over non-fiction writers. If the set would
grow too much in size, it would contain too many non-writers. On the
other hand, we think that it is important that the set contains most
famous writers, omitting only some special cases where the
classification of the particular person is ambiguous. 
%
%
%
The \apInfoboxesEn approach best meets our requirements, given that it
identifies the largest number of Nobel laureates (102) in literature
within a modest number of authors (25,995) and lacks only few
articles (633) that were identified as writers by the other
approaches.

Just in a few cases \emph{Writer} templates are also used for
articles on persons we would not usually classify as writers. We will
filter such cases by using a measure of importance
(cf. Section~\ref{sec:creation-dataset}).
On the other hand, we have to face the problem that not every article
on a famous writer necessarily contains a \wptype{Writer}
infobox. For example, Italian writer and philosopher \wpent{Umberto
  Eco} is not categorized as a writer referring to the
\wpfile{instance\_types\_en} dataset, since his article does not
contain a \wptype{Writer} infobox but an infobox of the type
\wptype{Philosopher}, just like in the cases of \emph{Homer} or
\emph{Albert Camus}. Others are also lacking a \wptype{Writer}
infobox, like \emph{Franz Kafka}, \emph{Geoffrey
  Chaucer} and \emph{Marcel Proust}, to name the most prominent
articles on writers (by number of their in-links) that feature a
\wptype{Person} infobox rather than one of the \wptype{Writer} type
(cf. also Table~\ref{tab:topfive_combined}). This is an issue of the
used datasets that leaves room for further work, but with the
exception of these admittedly major omissions we just listed plus
probably a handful more, we still catch a majority of what represents
world literature inside Wikipedia as our results will show.  On the
plus side, this approach ignores the many persons who are only
parenthetically active as writers, like, perhaps, a sportsman who
published his autobiography.
%

With respect to the analysis of language editions other than
English, we have to note that the \apInfoboxesEn approach omits
writers in other language versions that are neither contained nor
classified as writers in the English Wikipedia. We assume that the
English Wikipedia covers most famous writers and contributors to world
literature, whatever their writing languages may be (given that
English works as \emph{lingua franca} and taking into account that the
English version is by far the largest corpus within the Wikipedia
family). At a later stage, this approach can also be adopted if it
comes to analyzing literary works using the \wptype{WrittenWork} type
of the DBpedia ontology instead of the \wptype{Writer} type.
%
%

\section{Data Selection and Analysis}\label{sec:analysis}

In this section we first describe the creation of our basic set, which
consists only of writers represented by an article in the
English Wikipedia. We analyze the basic set in
Section~\ref{sec:analyzing-basic-set} and then extend our analysis to
14 other Wikipedia language versions.


\subsection{Creation of the Basic Set}\label{sec:creation-dataset}

As described in Section~\ref{sec:datasets}, we used the DBpedia 2014
download for our analysis. To build our \emph{basic set}, based only
on data from the English Wikipedia, we extracted all instances from
the \wpfile{instance\_types\_en} dataset that are of the type
\wptype{Writer}.
We extracted 25,995 instances, each of which represents an article in
the English Wikipedia.
%
For each instance we gathered information from different datasets on
DBpedia, \eg
we counted the number of incoming links (in-links) from within
the English Wikipedia pointing to every other article by analyzing
the \wpfile{page\_links\_en} dataset. The number of in-links can be
used as an indicator for the importance of an article. We also
added some infobox properties from the
\wpfile{mappingbased\_properties\_en} dataset, such as
\wpprop{birthDate}, \wpprop{deathDate}, and \wpprop{nationality}.
\newcommand{\plwidth}{.3\linewidth}%
\begin{figure*}
  \centering
  \begin{subfigure}{\plwidth}
    \plosincludegraphics[width=\linewidth]{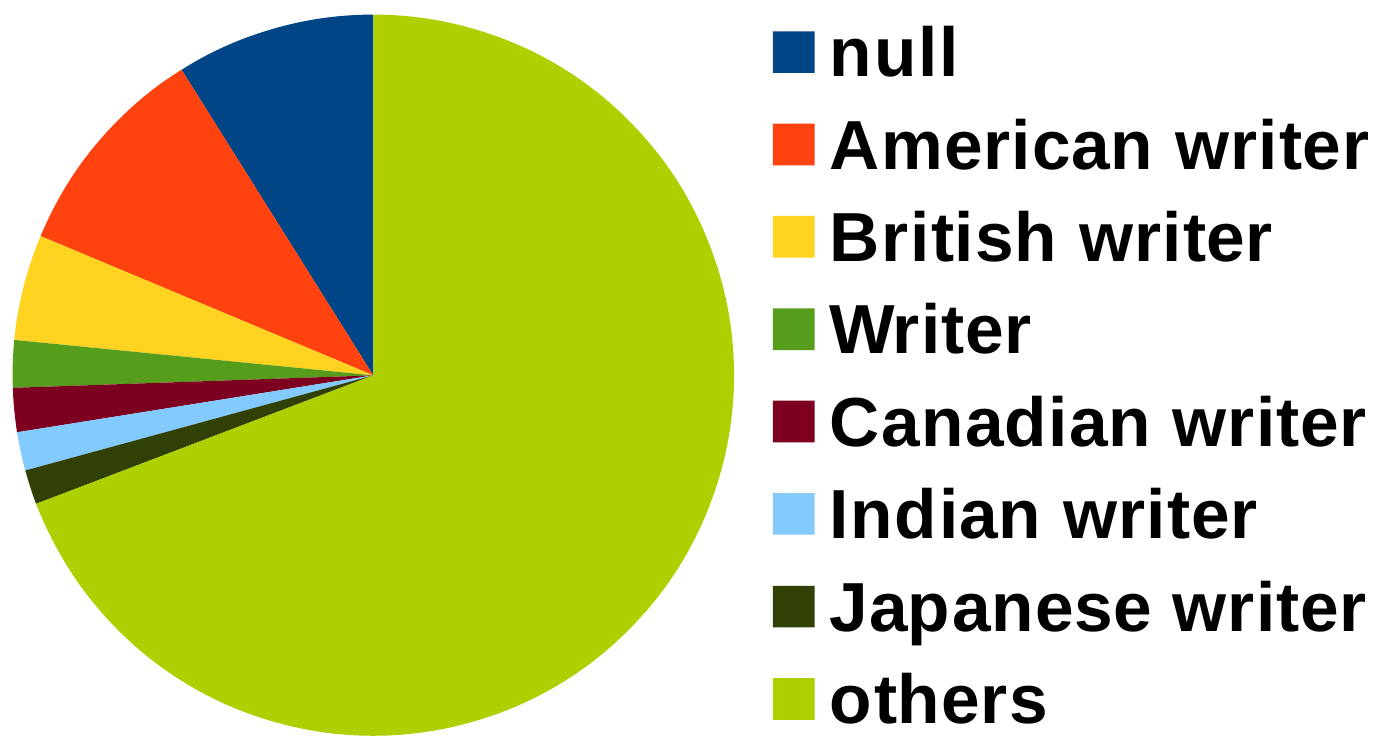}\subcaption{\wpprop{description}}
  \end{subfigure}\hfill
  \begin{subfigure}{\plwidth}
    \plosincludegraphics[width=\linewidth]{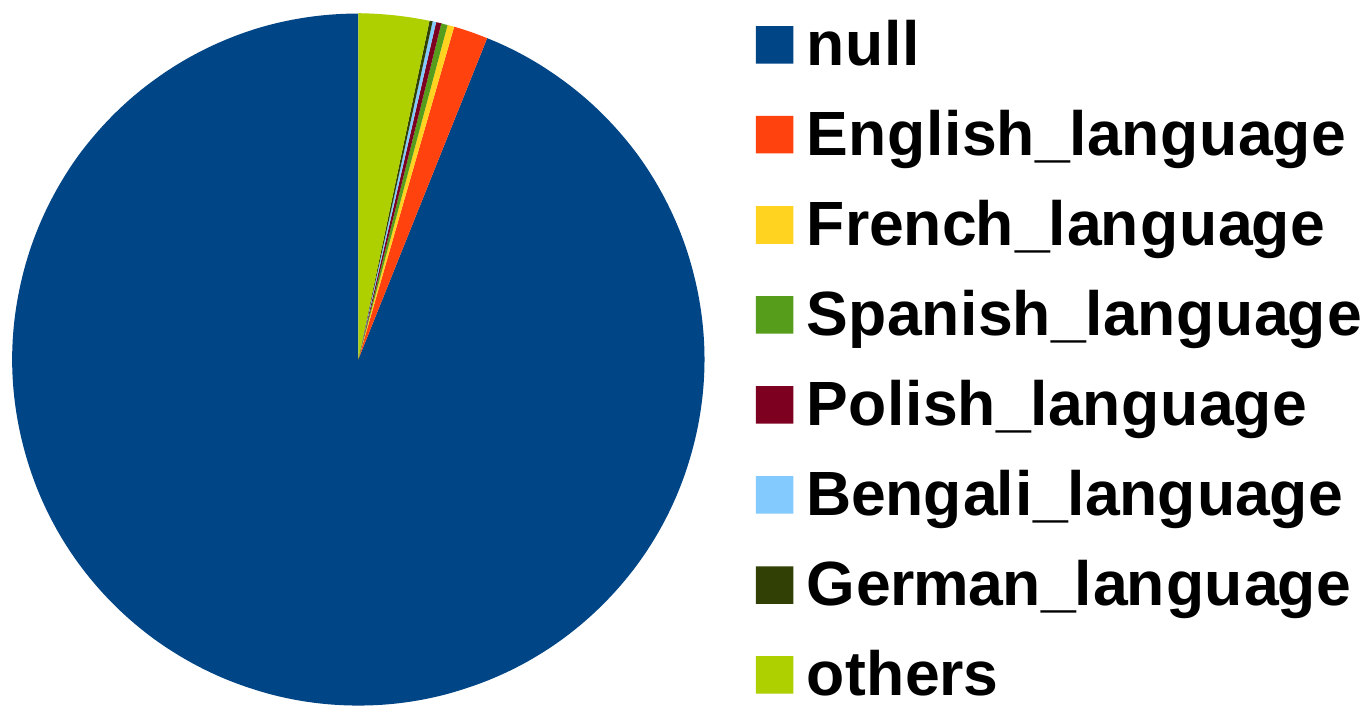}\subcaption{\wpprop{language}}
  \end{subfigure}\hfill
  \begin{subfigure}{\plwidth}
    \plosincludegraphics[width=\linewidth]{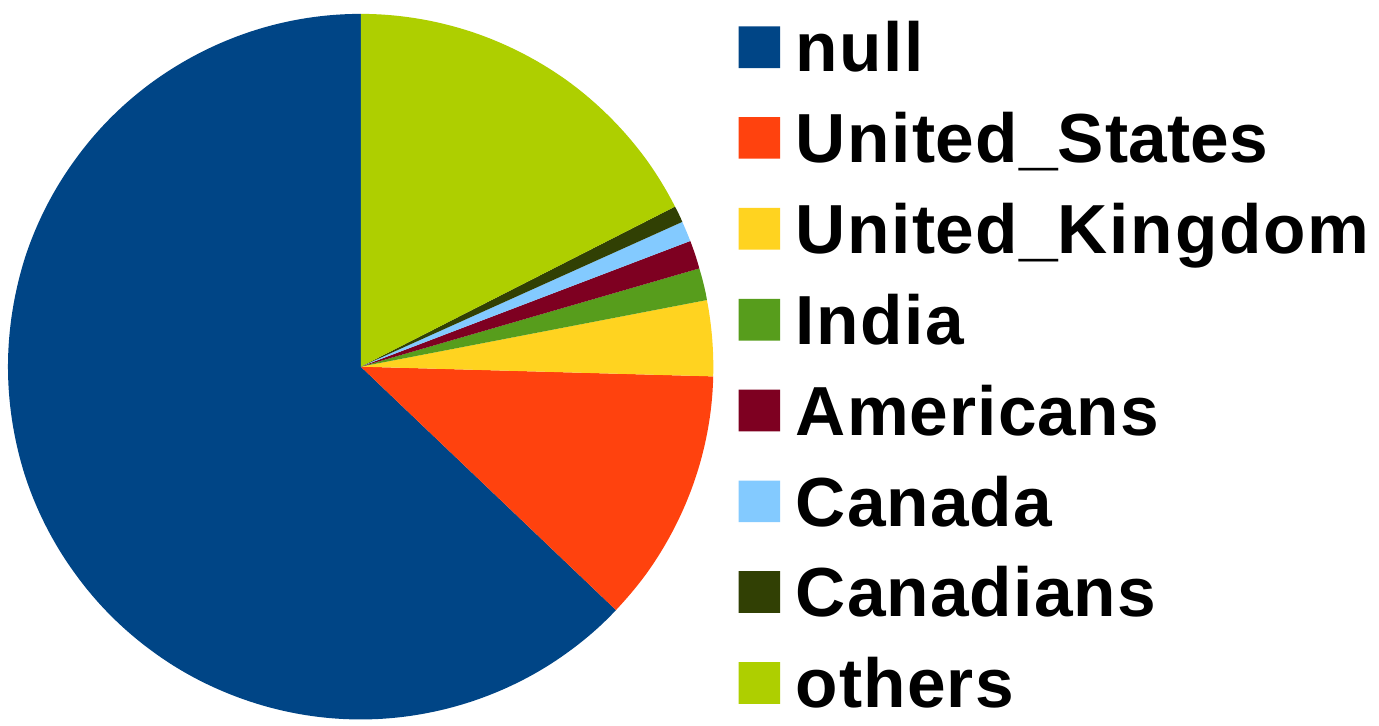}\subcaption{\wpprop{nationality}}
  \end{subfigure}
  \caption{Distribution of property values within infoboxes of
    articles of type \wptype{Writer} from the English
    Wikipedia.}\label{fig:properties}
\end{figure*}

%
Fig.~\ref{fig:properties} depicts the number of distinct terms used
for some properties, where the value \emph{null} indicates cases where
the property is not used in the infobox of an article at all.
The plots show that the \wpprop{description} property is frequently
used in \wptype{Writer} infoboxes and contains many diverse values,
while \wpprop{language} and \wpprop{nationality} are used only
marginally.
Given these distributions, it is difficult to identify the languages
of writers from the data provided by the infoboxes. 
(Although
  a \wpprop{description} is frequently provided, it typically only
  contains the nationality of the writers which, of course, does not
  necessarily indicate their writing language(s).) 
%
Therefore, if we want to analyze writers in consideration of their
writing languages, we have to find alternative methods (something we
will undertake in Section~\ref{sec:native-writers}).
%
%
In general, most infoboxes apply only a few properties, with
a few exceptions such as \wpprop{birthDate} and \wpprop{name}.
The \wpprop{birthDate} property is used by 21,607 of 25,995
instances. Other instances only involve the \wpprop{birthYear} property
or no birth-related property at all.

Some articles on lesser known persons use the \wptype{Writer} template out
of context, probably in lack of a better fitting person template.
Aiming at a higher precision, we
decided to delete instances that neither contain a value for
\wpprop{birthDate} or \wpprop{birthYear} nor for \wpprop{deathDate} or
\wpprop{deathYear}, assuming that for every important writer at least
one of these properties would have been added to the infobox of the
corresponding article (an approximate date would be enough).
%
%
Additionally, we removed instances with less than 10 incoming links,
using this number as a measure of importance. We hope that by this
approach we can also evade most bot-generated articles.
Eventually, our basic set includes 10,765 writers.
As we will see in Section~\ref{sec:nobel-laureates}, this filtering
step did not remove any Nobel laureates in literature, therefore, the
resulting set of writers has the highest precision 
and recall 
compared to the results in Table~\ref{tab:topfive_combined}.
%


\subsection{Analyzing the Basic Set}\label{sec:analyzing-basic-set}

As mentioned in Section~\ref{sec:choosing_approach},
%
%
the English Wikipedia takes on a special role within the plenitude of
Wikipedia language versions. Not just because English acts as a
\emph{lingua franca}, but also because the English Wikipedia is by far
the largest of all language versions and we also expect it to
explicate a greater cultural diversity than most other language
versions.

\begin{figure}
  \centering
  \plosincludegraphics[width=.7\linewidth]{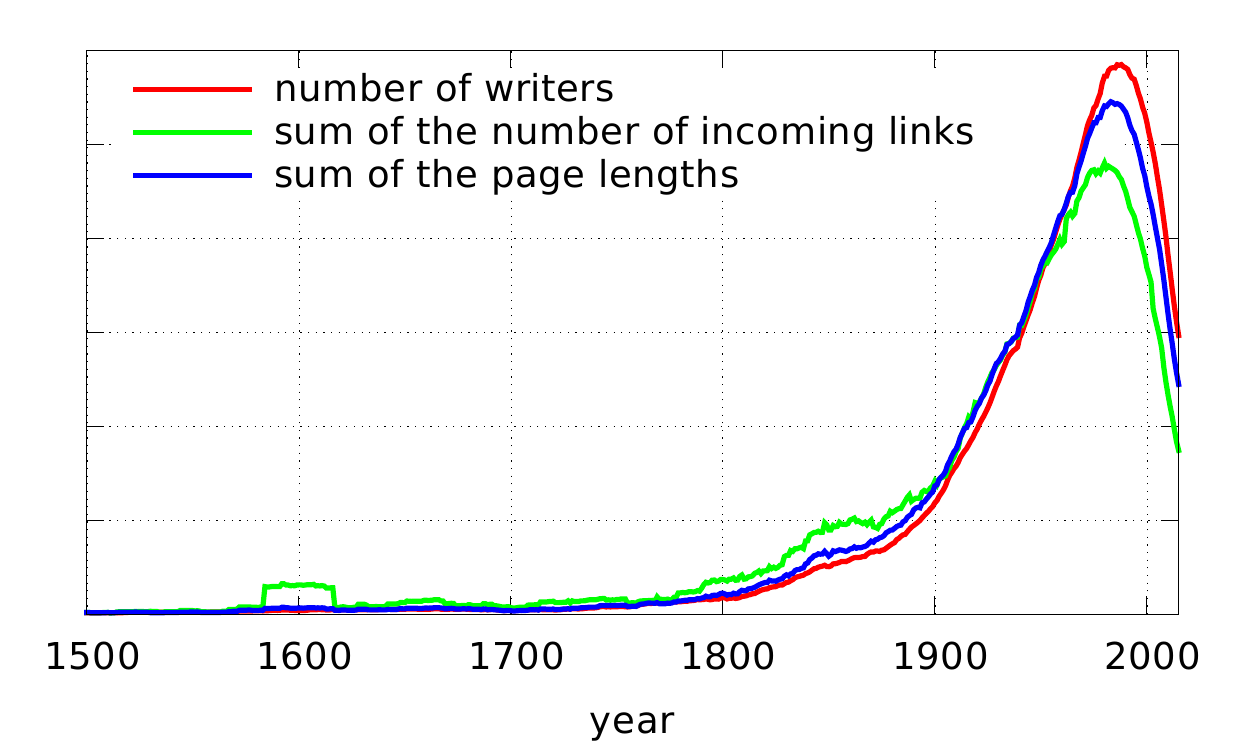}
  \caption{(Normalized) number of writers, sum of their articles'
    page lengths, and sum of their articles' number of incoming links
    from 1500 to 2015 in the English
    Wikipedia.}\label{fig:numbersEnglish}
\end{figure}

To gain a first impression of the distribution of writers over time,
we approximate their time-of-writing activity from the age of 20 until
the age of 60 (unless they died earlier, of course).
For instance, for a writer who was born in 1820 and died in 1890, we
would assume an active phase from 1840 to 1880.
We are aware that this is a rather non-realistic approximation, but it
helps handle the fact that some writers in the set have no death
date, and it will still show us epochal peaks.
%
%
%
For each year we now count the number of writers that were `active'
that year, the sum of incoming links their articles received, and the
sum of the page lengths of their articles. To make these three values
comparable, we normalized them so that each area under the curve is
equal to one.
The resulting distribution for all writers active since 1500 is shown
in Fig.~\ref{fig:numbersEnglish} and is based on the data provided by
our basic set for the English Wikipedia. 
(Only very few writers are contained from before 1500.) 
We omitted labels for the
$y$-axis, since we want to focus on the order of magnitude of the
curves.

As we can see, the numbers show a significant rise after 1800 and
increase until around 1990. The reason for this can simply be
attributed to the fact that Wikipedia's coverage of recent events is
much broader, as has been indicated by \cite{denning2005risks} and
\cite{Elvebakk2008Philosophy}, among others.  (Writers who were born
after 1970, though, are less well represented as yet.)  But we are
also reminded of the division of world literature into an `old' and a
`new' type, as suggested by \cite{levine2013what}, arguing that there
is a `timeless' and `deeply historical' world literature and an
`ephemeral', `newly emerging' one.

The three distributions in Fig.~\ref{fig:numbersEnglish} are very
similar. The noticeable rise in the sum of incoming links between 1584
and 1616 can be attributed to the immense influence of William
Shakespeare who to no surprise is ``widely regarded as the greatest
writer in the English language and the world's pre-eminent
dramatist'', according to the English Wikipedia
itself \cite{wp:shakespeare}. 
Shakespeare also benefits from the fact that his epoch is represented
by only a few writers, which is why his influence can be easily noted
in the graph. Famous modern writers are hard to perceive in the diagram
due to the large number of competing writers in the set.
%



\subsection{Other Language Versions}\label{sec:other-lang-vers}

As we have seen in Section~\ref{sec:choosing_approach}, explicit
information about writers is not available in some language
versions. Furthermore, for only a small fraction of writers in our
basic set their language is explicitly specified in Wikipedia. It
is thus difficult to analyze important writers by their writing
language. Hence, in this section we present an approach to
identify writers in other language versions of Wikipedia and analyze
their distribution over time.

We extracted writers from each language version by using the
\wpfile{interlanguage\_links\_en} dataset, which maps each article
from the English Wikipedia to the corresponding articles in other
language versions if such a corresponding article exists.  Obviously,
not every writer from our basic set can be found in each language
version. The column ``\#writers in basic set'' in
Table~\ref{tab:languages} shows how many of the 10,765 writers of our
basic set can be found in each Wikipedia language version. It is
important to understand that only the writers of our basic set have
been taken into account, so writers who are not represented in the
English Wikipedia but in other language versions were not
considered.
The writers also do not necessarily need to be classified
as \wptype{Writers} in the other 
versions as long as they are represented by an article.
This approach, as described in Section
\ref{sec:writer_infoboxes_english}, helped us to bypass
the problem that \wpfile{instance\_types} datasets are only available
for a small number of language versions. At the same time, the
values obtained in this way have to be interpreted cautiously
so as not to simply reinscribe the English model of world literary
history as \emph{the} model of world literature.





\begin{figure}
  \centering
    \plosincludegraphics[width=.7\linewidth]{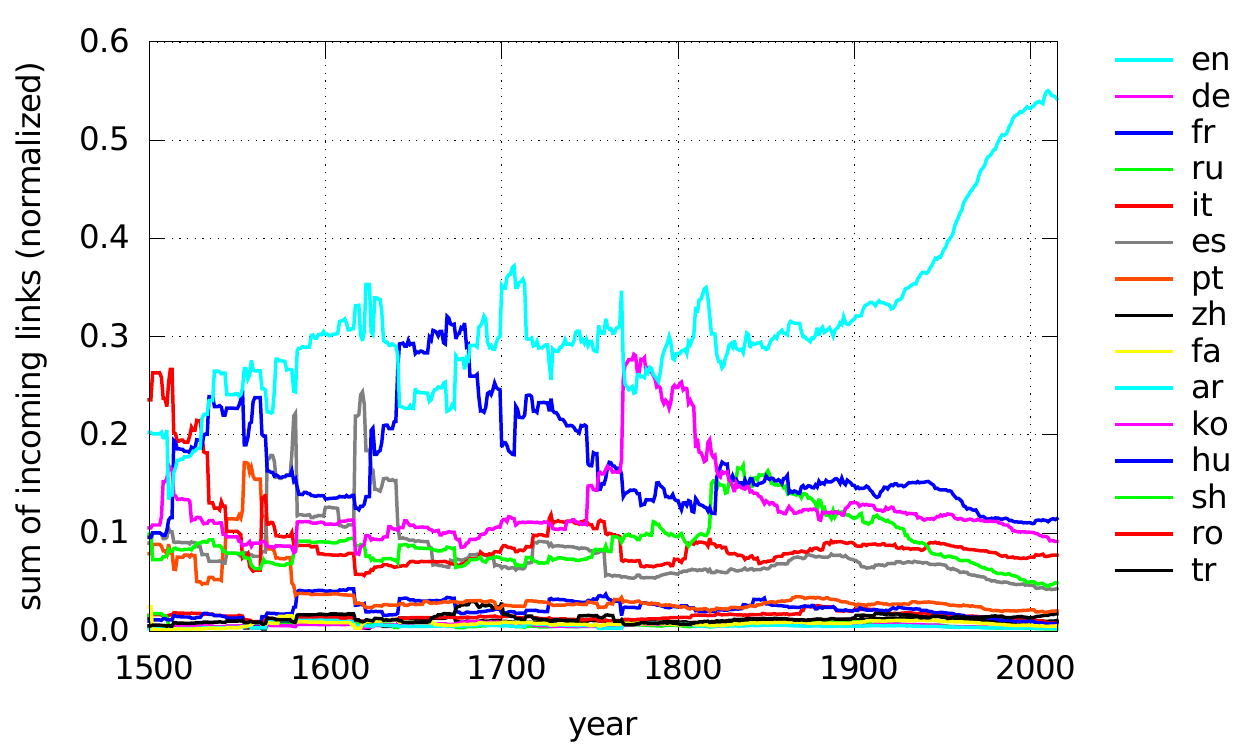}
    \caption{Distribution of incoming links for writer articles in
      different language versions.}\label{fig:inlinks}
\end{figure}


Like we showed before, a glance at the number of incoming links over
time in Fig.~\ref{fig:numbersEnglish} as a measure of relevance
allowed us to identify the importance of \wpent{Shakespeare} for the
English Wikipedia.  Accordingly, we now analyze this distribution over
all 15 language versions.
We assume that writers who have been linked to frequently within a
Wikipedia language version play an important role in this version.
Fig.~\ref{fig:inlinks} depicts the sum of in-links of active writers
by time in a normalized manner to neutralize the exponential rise of
the number of writers in the modern age (meaning that the sum of the
values for all language versions add up to one for each year).


It comes as no surprise that the English Wikipedia dominates the
scene, especially regarding the 20th century, since all other language
version sets of our approach are subsets of the English set.
However, we can recognize several periods of time where the curves for
other language versions stand out. It does not seem too far-fetched
that these time periods correlate with major literature epochs in the
particular languages. For instance, the number of incoming links of
active writers in the German Wikipedia stands out, roughly, between
the last third of the 18th century and the first third of the 19th
century.  These decades see the rise of the \emph{Sturm und Drang}
movement and the emergence of German Romanticism and Classicism,
associated with authors such as \wpent{Johann Wolfgang von Goethe} or
\wpent{Friedrich Schiller}.  The influence of Portugal's most famous
poet and national hero \wpent{Luís de Camões} leads to a peak for the
Portuguese Wikipedia at around 1550.  The highest peak in the French
Wikipedia pertains to the last two-thirds of the 17th century, the era
of French Classicism under the reign of Louis XIV when playwrights
like \wpent{Pierre Corneille}, \wpent{Jean Racine}, and
\wpent{Molière} ruled the scene.  Russian literature culminates in the
middle of the 19th century which coincides with the life spans of
\wpent{Alexander Pushkin}, \wpent{Nikolai Gogol}, and \wpent{Fyodor
  Dostoyevsky}.  The Spanish curve peaks around 1600, the life spans
of \wpent{Cervantes} and \wpent{Lope de Vega}.  The harsh recess in
the years before and after 1600 is a result of our normalization
and the strong influence of \wpent{Shakespeare} around that time. The
curve starts to rise again right after his death in 1616 and falls
sharply only after the death of \wpent{Lope de Vega} in 1635.

\enlargethispage{\baselineskip}

%
As already mentioned in Section~\ref{sec:analyzing-basic-set}, it is
much easier to identify writers and their influences before 1900
because of the strongly increasing number of writers in the modern and
postmodern age.

The conclusions that can be drawn from Fig.~\ref{fig:inlinks} are even
more remarkable if we recap that for each language version we only
consider persons who are flagged as \wpent{Writer} in
the \emph{English} Wikipedia. E.g., although the coverage of English
authors presumably reaches its optimum in the English Wikipedia, our
approach does not automatically ensure a high coverage of German authors in the
German Wikipedia. Nevertheless, just the distribution of incoming
links over time allows us to identify the corresponding curve as that
of the German Wikipedia. Since this claim also holds for other major
language editions, this means that the language editions have a strong
focus on their ``own'' writers (as already indicated by
\cite{Callahan2011culturalbias}) and thereby allow us to identify
important literary epochs for each language, despite of considering
only authors that are also contained in the English language
edition. But for the time being, the unevenness of the per-language
datasets should hold us back from reading too much into our
experimental setup regarding other language versions.




\section{Most Prominent Writers in Wikipedia}\label{sec:most-import-writ}

After all this preliminary work, let us come back to our initial
purpose, the question as to how world literature is represented on
Wikipedia. On the one hand, this means we want to get a fresh
look at world literature without necessarily comparing the results to
well-established canons. On the other hand, we propose several methods
for ranking writers on Wikipedia, and it is difficult to judge which
method can best be used to identify important writers without a
comparison to some ground truth.
To get out of this dilemma, we perform three different types of analyses:
\begin{enumerate}
\item We compare the different rankings by computing correlations
  for pairs of rankings. This allows us to identify similar rankings.
\item We evaluate how well the measures rank the Nobel laureates in
  literature to the top positions. This gives us an estimate on how
  a very specific canon of writers is represented on Wikipedia.
\item We analyze the graph of top writers in the English Wikipedia
  formed by the links between their articles. This provides
  insights into the link structure and potential groups of
  well-connected writers.
\end{enumerate}
In addition, Section~\ref{sec:native-writers} is dedicated to writers
and their writing languages and is looking at the performance of
authors in Wikipedia versions whose language is different from the
writing language of the author.
But let us start this section by motivating and explaining the
proposed ranking approaches before conducting the above-mentioned
analyses.

\subsection{Ranking Measures}\label{sec:ranking-measures}

Five different ranking measures are considered, among them basic
measures that count properties of the writers' article pages (e.g.,
their page length) and more complex ones based on the PageRank
algorithm \cite{brin1998anatomy}:
%
\begin{description}

\item[page length (\rpagelen):] We rank writers by the \emph{page
    length} of their articles, as provided by the
  \wpfile{page\_length} datasets.

\item[number of in-links (\rinlinks):] As we have seen, the
  \emph{number of in-links} and their temporal distribution
  allow us to distinguish between different
  language versions. This value is also more robust against
  manipulation than the page length, since it requires changes to many
  pages,
  and it is clearly an indicator of relevance if a Wikipedia page
  is linked to from many other pages.
%
%

\item[PageRank writers (\rprwriter):] We additionally consider the
  importance of writers for other writers by computing the
  \emph{PageRank} \cite{brin1998anatomy} of each article on the link
  graph of all the \emph{writers} contained in our sets. PageRank
  is widely accepted as a solid measure for the relevance of nodes
  within a graph.
  A writer has a high PageRank if many writer articles with a high
  PageRank link to it.
  %
\item[PageRank complete (\rprcomplete):] We also compute the
  \emph{PageRank} for the \emph{complete} English Wikipedia and then
  extract the ranking for writers from our basic set.  This ranking
  can be regarded as an indicator for the importance of writers among
  all Wikipedia articles.
  %

\item[number of page views (\rpvtwo, \rpvthree, \rpvfour):] The
  previous measures reflect properties of Wikipedia itself, i.e., as
  it was created by its editors. We supplement these with a measure
  that captures the importance of articles according to the visitors
  of Wikipedia. For that, we use the Wikipedia page view dataset
  \cite{pagecounts}. 
  We extract information on how often a writer's article page was
  accessed in 2012, 2013, and 2014 (corresponding to measure \rpvtwo,
  \rpvthree, and \rpvfour, respectively). We also include accesses to
  redirect pages in those counts, e.g., the page
  \url{https://en.wikipedia.org/wiki/Goethe} redirects to
  \url{https://en.wikipedia.org/wiki/Johann_Wolfgang_von_Goethe} --
  accesses to both pages increment the count for \emph{Johann Wolfgang
    von Goethe}.
\end{description}



The English Wikipedia's top five writers for each approach are shown
in Table~\ref{tab:topfive}. We also show the results for the other 14
language editions for the approaches \emph{number of in-links}
(\rinlinks) and \emph{PageRank writers} (\rprwriter).
The top 25 writers for all seven measures can be found on
\url{http://data.weltliteratur.net/}.
%

%
%
%
\newcommand{\n}[1]{\sffamily \textbf{#1}}


\begin{sidewaystable}
  \caption{\bf Top five writers according to different ranking approaches.}\label{tab:topfive}
  \footnotesize
  \setlength{\tabcolsep}{2pt}
  \begin{tabular}{@{}lllllllrr@{}}
    & Wikipedia      & 1st                    & 2nd                   & 3rd                     & 4th                    & 5th & \multicolumn{2}{l}{count\textsubscript{25}}\\
     \midrule
     %
     %
     \rpagelen 
     &English        &Mircea Eliade            &Mihail Sadoveanu        &Ion Luca Caragiale      &Benjamin Fondane     &Alexandru Macedonski & 14&(56\%) \\
     \midrule
     \rinlinks 
     &English        & \n{W. Shakespeare}        & \n{Robert Christgau}  & \n{Roger Ebert}           & \n{Charles Dickens}       & \n{J.\,R.\,R. Tolkien}  &18 &(72\%)\\
     &German         & \n{J.\,W. von Goethe}     & W. Shakespeare        & \n{Friedrich Schiller}    & \n{Bertolt Brecht}        & \n{Thomas Mann}         &11 &(44\%)\\
     &French         & W. Shakespeare            & \n{Victor Hugo}       & \n{Molière}               & \n{Voltaire}              & Anton Chekhov           &15 &(60\%)\\
     &Russian        & \n{Alexander Pushkin}     & W. Shakespeare        & \n{Anton Chekhov}         & \n{Maxim Gorky}           & \n{Leo Tolstoy}         &13 &(52\%)\\
     &Italian        & \n{Dante Alighieri}       & W. Shakespeare        & \n{Cicero}                & J.\,R.\,R. Tolkien        & \n{Virgil}              &14 &(56\%)\\
     &Spanish        & W. Shakespeare            & J.\,R.\,R. Tolkien    & \n{Miguel de Cervantes}   & \n{Lope de Vega}          & Cicero                  & 9 &(36\%) \\
     &Portuguese     & W. Shakespeare            & Robert Christgau      & J.\,R.\,R. Tolkien        & \n{Vinicius de Moraes}    & Cicero                  & 9 &(36\%) \\
     &Chinese        & \n{Lu Xun}                & J.\,R.\,R. Tolkien    & W. Shakespeare            & \n{Jin Yong}              & \n{Li Bai}              &12 &(48\%)\\
     &Persian        & \n{Ferdowsi}              & \n{Ahmad Shamloo}     & W. Shakespeare            & \n{Ali-Akbar Dehkhoda}    & \n{Mohammad-Taqi Bahar} &10 &(40\%)\\ 
     &Arabic         & W. Shakespeare            & \n{Naguib Mahfouz}    & J.\,W. von Goethe         & \n{Al-Maqrizi}            & \n{Ahmed Shawqi}        & 9 &(36\%) \\
     &Korean         & W. Shakespeare            & \n{Yi Kwang-su}       & J.\,W. von Goethe         & J.\,R.\,R. Tolkien        & Dante Alighieri         & 4 &(16\%) \\
     &Hungarian      & W. Shakespeare            & \n{Sándor Petőfi}     & \n{János Arany}           & \n{Attila József}         & Robert Christgau        & 7 &(28\%) \\
     &Serbo-Croatian & W. Shakespeare            & \n{Vjekoslav Klaić}   & Cicero                    & Roger Ebert               & Plutarch                & 3 &(12\%) \\
     &Romanian       & \n{Ion Luca Caragiale}    & \n{Mihai Eminescu}    & W. Shakespeare            & Jules Verne               & \n{Mircea Eliade}       &10 &(40\%)\\
     &Turkish        & W. Shakespeare            & Bertolt Brecht        & J.\,R.\,R. Tolkien        & Anton Chekhov             & Molière                 & 7 &(28\%) \\
     \midrule
     \rprwriter 
     &English        & \n{W. Shakespeare}        & \n{T.\,S. Eliot}      & \n{Charles Dickens}       & J.\,W. von Goethe         & \n{Ernest Hemingway}   &18 &(72\%)\\
     &German         & W. Shakespeare            & \n{J.\,W. von Goethe} & Voltaire                  & \n{Thomas Mann}           & \n{Friedrich Schiller} & 7 &(28\%) \\
     &French         & W. Shakespeare            & \n{Victor Hugo}       & \n{André Gide}            & J.\,W. von Goethe         & \n{Charles Baudelaire} &14 &(56\%)\\
     &Russian        & Heinrich Heine            & \n{Alexander Pushkin} & Alexander Pope            & W. Shakespeare            & \n{Gavrila Derzhavin}  & 9 &(36\%) \\
     &Italian        & W. Shakespeare            & \n{Dante Alighieri}   & J.\,W. von Goethe         & \n{Virgil}                & \n{Petrarch}           & 8 &(32\%) \\
     &Spanish        & W. Shakespeare            & Lord Byron            & Dante Alighieri           & \n{Jorge Luis Borges}     & Edgar Allan Poe        & 2 &(8\%)  \\
     &Portuguese     & W. Shakespeare            & Stephen King          & T. S. Eliot               & F. Scott Fitzgerald       & Zelda Fitzgerald       & 0 &(0\%)  \\
     &Chinese        & W. Shakespeare            & Victor Hugo           & Dante Alighieri           & George Orwell             & \n{Lu Xun}             & 1 &(4\%)  \\
     &Persian        & \n{Sanai}                 & W. Shakespeare        & Alexander Pushkin         & Nikolai Gogol             & \n{Ferdowsi}           & 5 &(20\%) \\
     &Arabic         & W. Shakespeare            & Christopher Marlowe   & Leo Tolstoy               & Friedrich Schiller        & Fyodor Dostoyevsky     & 1 &(4\%)  \\
     &Korean         & W. Shakespeare            & Dante Alighieri       & J.\,W. von Goethe         & Friedrich Schiller        & Virgil                 & 0 &(0\%)  \\
     &Hungarian      & W. Shakespeare            & Giovanni Boccaccio    & Petrarch                  & J.\,W. von Goethe         & André Breton           & 3 &(12\%) \\
     &Serbo-Croatian & W. Shakespeare            & Fyodor Dostoyevsky    & Charles Dickens           & Virgil                    & Dante Alighieri        & 0 &(0\%)  \\
     &Romanian       & W. Shakespeare            & J.\,W. von Goethe     & Miguel de Cervantes       & Charles Baudelaire        & Molière                & 3 &(12\%) \\
     &Turkish        & W. Shakespeare            & \n{Peyami Safa}       & J.\,R.\,R. Tolkien        & C.\,S. Lewis              & Charles Baudelaire     & 2 &(8\%)  \\
     \midrule
     \rprcomplete
     &English        &\n{W. Shakespeare}     & Cicero                  & \n{Charles Dickens}    &    J.\,W. von Goethe   &  \n{J.\,R.\,R. Tolkien} & 17 & (68\%) \\
     \midrule
     \rpvtwo
     &English  &\n{W. Shakespeare} & \n{Ernest Hemingway} & \n{Charles Dickens} & \n{Edgar Allan Poe} & \n{Dr. Seuss} & 23 & (92\%)\\
     \midrule
     \rpvthree
     &English & \n{W. Shakespeare} & \n{Ernest Hemingway} & \n{Edgar Allan Poe} & \n{Douglas Adams} & \n{J.\,R.\,R. Tolkien} & 23 & (92\%)\\
     \midrule
     \rpvfour
     &English & \n{W. Shakespeare} & \n{Maya Angelou} & \n{Ernest Hemingway} & \n{J.\,K. Rowling} & \n{George R.\,R. Martin} & 21 & (84\%)\\

  \end{tabular}
\end{sidewaystable}



This tool can be put to use in several ways. First of all, it makes it easy
to see how the hypercanon of world literature is represented across languages,
but also how there are slight differences between the presence and rank. But
just as important, one should be observant of more than just the
hypercanonical strata in world literature and bring more attention to the
shadow canon and the countercanon, as David Damrosch suggested in
``World Literature in a Postcanonical, Hypercanonical Age''
\cite{damrosch2006postcanonical}. While Damrosch acknowledges that the
strength of hypercanonical writers seems to grow in terms of critical work,
e.g., the dominance of a few writers in British romanticism or the attention
brought to Salman Rushdie’s work in postcolonial studies, he calls for an
awareness of writers that may represent different literary qualities and
norms that distinguish them from the hypercanonical authors, which could
be called the countercanon. The dynamics of the literary field also
produces a shadow canon of works whose canonical status used to be
undisputed but which are now showing to be more in peril of being
forgotten. The data gathered at \url{http://data.weltliteratur.net/} makes it
easy to observe these dynamics that are grounded in the multifaceted use
of Wikipedia and demonstrates, for example, the countercanonical
fascination of science fiction or the endurance of writers such as Ernest
Hemingway and Charles Bukowski in other languages.



\subsection{Ranking Correlation}\label{sec:ranking-correlation}

Let us first answer the question of how similar the ranking measures
are to each other.
From a mathematical point of view, some measures are closely related,
i.e., the two \emph{PageRank} variants and the \emph{number of
  in-links}, since all three of them are based on the link structure
between articles.
Therefore, we expect a high correlation among the corresponding
rankings.
It is also interesting to check whether some of the intrinsic
measures are correlated to the extrinsic \emph{page views} measures.

\enlargethispage{\baselineskip}

For this analysis we compute the two rank correlation coefficients
Spearman's $\rho$ and Kendall's $\tau$, which are standard measures for
the comparison of rankings.
For each pair of rankings we compute the two correlation coefficients
for each language edition. For each such pair we then calculate the
mean over all language editions.
%
The resulting correlation coefficients for all pairs of ranking
measures can be found in Table~\ref{tab:correlations}.
%
%
%
\begin{table}\centering
  \caption{\bf Rank correlation coefficients between different ranking measures.}
  \label{tab:correlations}
  
  \begin{tabular}{@{}lrrrrrrr@{}}
    \toprule

                 &\rpagelen &\rinlinks &\rprwriter &\rprcomplete &\rpvtwo &\rpvthree &\rpvfour \\
    \rpagelen    &          &0.541     &0.388      &0.496        &0.545   &0.574     &0.583\\    
    \rinlinks    &0.391     &          &0.606      &0.886        &0.765   &0.768     &0.773\\    
    \rprwriter   &0.291     &0.478     &           &0.558        &0.504   &0.496     &0.497\\    
    \rprcomplete &0.349     &0.731     &0.429      &             &0.744   &0.735     &0.743\\    
    \rpvtwo      &0.390     &0.594     &0.386      &0.560        &        &0.947     &0.900\\    
    \rpvthree    &0.409     &0.595     &0.378      &0.550        &0.849   &          &0.953\\    
    \rpvfour     &0.415     &0.601     &0.379      &0.557        &0.777   &0.847     &\\         
    \bottomrule
  \end{tabular}
  \begin{flushleft}
    The upper triangular matrix shows Spearman's~$\rho$, the lower
    triangular matrix Kendall's~$\tau$.
  \end{flushleft}
\end{table}
%
%

%
%
We can observe a high consistency between Kendall's $\tau$ and
Spearman's $\rho$, with Kendall's $\tau$ showing an overall lower
correlation than Spearman's $\rho$. 
In general, all pairs of rankings are positively correlated, most of
them show a medium-to-high correlation. This indicates that the
measures are similar to a certain extent.
Most highly correlated to each other are the \emph{page views}
\rpvtwo, \rpvthree, \rpvfour\ -- consistently for both correlation
measures. This shows that the interest of the users of Wikipedia in
certain writers remains fairly constant over time but is also subject
to slight changes, since the page views of subsequent years have a
higher correlation than the page views of \rpvtwo and \rpvfour.
Aside from the page views, the \emph{number of in-links} (\rinlinks)
and the \emph{PageRank complete} (\rprcomplete) are most highly
correlated which supports our expectation and indicates their close
relationship.
\rinlinks and \rprcomplete are also most highly correlated to the page
views, which means that those two intrinsic rankings best reflect the
extrinsic rankings.
The lowest correlations to all other rankings are exhibited by the
\emph{page lengths} (\rpagelen) and the \emph{PageRank writers}
(\rprwriter).
This is not too surprising with regard to page lengths, since this measure
can easily be influenced by the enthusiasm of individual Wikipedia
editors. We will observe this exemplarily when analyzing the
representation of native writers in
Section~\ref{sec:writers-their-own}.
\rprwriter shows the lowest correlation of all rankings to the page
views, which is quite surprising, given that the PageRank (although
computed on the complete web graph) still is one of the main
ingredients of most search engines' rankings which also direct users
to Wikipedia and therefore have a high influence on the number of page
views.
%
%
%
Overall, the results indicate that the \emph{page length} and the
\emph{PageRank complete} differ most from each other and from the
other three measures which in turn  are more similar to each other.


\subsection{Nobel Laureates}\label{sec:nobel-laureates}

One simple way to verify the ranking methods was to check how high
they ranked Nobel Prize winners in Literature, assuming that this
group of people qualifies as some kind of fallback canon.
%
%
We assembled the list of the 111 Nobel laureates so far, from 1901 to
2014, off of
Wikipedia \cite{wp:nobel}, 
and aligned it with the list of writers in our basic set.
As already observed in Section~\ref{sec:choosing_approach}, not all
laureates could be identified by the approach since nine persons are
missing in the basic set:
\emph{Theodor Mommsen} (1902), 
\emph{Rudolf Eucken} (1908), 
\emph{W.\,B. Yeats} (1923), 
\emph{Henri Bergson} (1927), 
\emph{Bertrand Russell} (1950), 
\emph{Winston Churchill} (1953), 
\emph{Albert Camus} (1957),
\emph{Jean-Paul Sartre} (1964), and 
\emph{Patrick Modiano} (2014).
%
%
This does not come unexpectedly, given that most of them are rather
historiographers or philosophers than novelists, playwrights, or
poets, so Wikipedians did not equip them with a \emph{Writer}
template.  \emph{W.\,B.~Yeats}, by the way, is not contained in the
basic set because the template in question was only added to his
article on May 19, 2015, which was before the time our DBpedia dump
was created.  On the whole, 102 of the 111 laureates are contained in
our basic set, which is quite good compared to other approaches
(cf. Table~\ref{tab:topfive_combined}).

To analyze which of the rankings has the Nobel laureates appear at the
highest positions, we use standard methods from machine learning --
the so-called \emph{ROC curve} (ROC = receiver operating
characteristic) and the \emph{AUC} (area under the curve)
\cite{fawcett2006introduction}. For each writer in the ranking
(starting at the top) we evaluate whether the writer is a Nobel
laureate in literature (\emph{true positive}) or not (\emph{false
  positive}). We use this information to draw the ROC curve by
starting at the coordinate position $(0,0)$ and then for each false
positive we go one step to the right (along the $x$-axis which
represents the \emph{false positive rate}) and for each true positive
one step to the top (along the $y$-axis which represents the
\emph{true positive rate}). 
  The step size for the $x$-axis
  ($y$-axis) is the reciprocal of the number of false (true)
  positives. 
\begin{figure}
  \centering
  \plosincludegraphics[width=.9\linewidth]{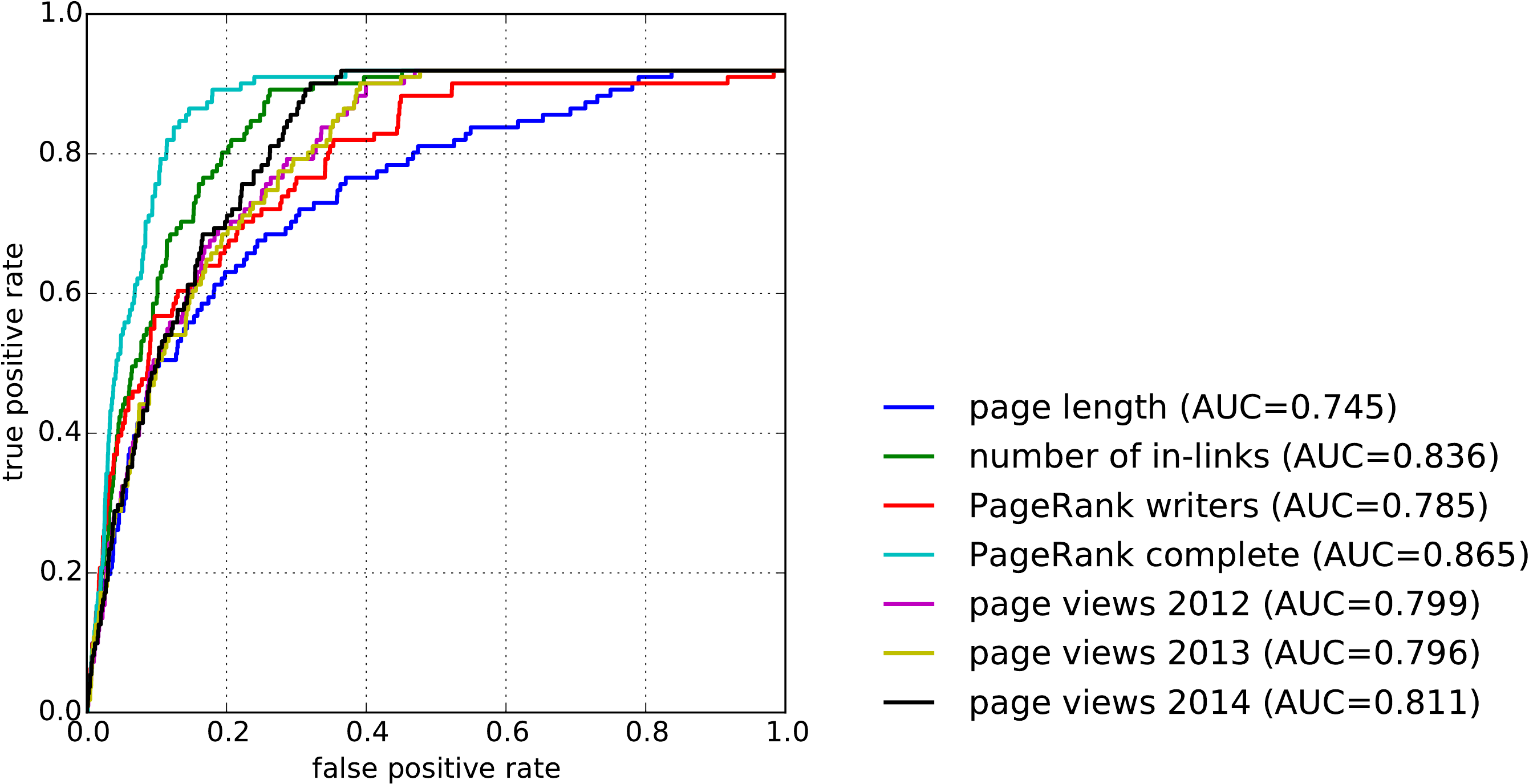}
  \caption{ROC curves for different approaches show how far to the top Nobel laureates are ranked.}
  \label{fig:roc}
\end{figure}
The resulting curves for each ranking measure can be seen in
Fig.~\ref{fig:roc}.
If all 111 Nobel laureates would appear in positions 1 to 111 (in any
order), the curve would go straight from $(0,0)$ to $(0,1)$ (and
then to $(1,1)$), resulting in
an AUC of 1.0. A random ranking would result in a straight line from
$(0,0)$ to $(1,1)$ with an AUC of 0.5.
The distance on the $y$-axis missing to 1.0 arises from the fact that
only 102 of 111 laureates appear in our basic set.

As we can see, the curve for the \emph{PageRank complete} stays closer
to the $y$-axis and therefore achieves the highest AUC of 0.865,
followed by the \emph{number of in-links} with an AUC of 0.863. This
means that Nobel laureates have a high PageRank and a high number of
incoming links in the English Wikipedia which clearly shows their
importance. We can also observe that the \emph{page length} has the
lowest AUC which indicates that, although the articles on the
laureates are well-linked in Wikipedia, they are often not as
comprehensive as other writers' articles. The \emph{page views} have
an AUC of around 0.8 which is lower than the best AUC and higher than
the worst. Articles on Nobel laureates are thus also viewed fairly
frequently by readers of Wikipedia.



\subsection{Writer Graph}\label{sec:writer-influences}

%

\begin{figure}
  \centering
  \makebox[\linewidth][c]{\plosincludegraphics[width=1.1\linewidth]{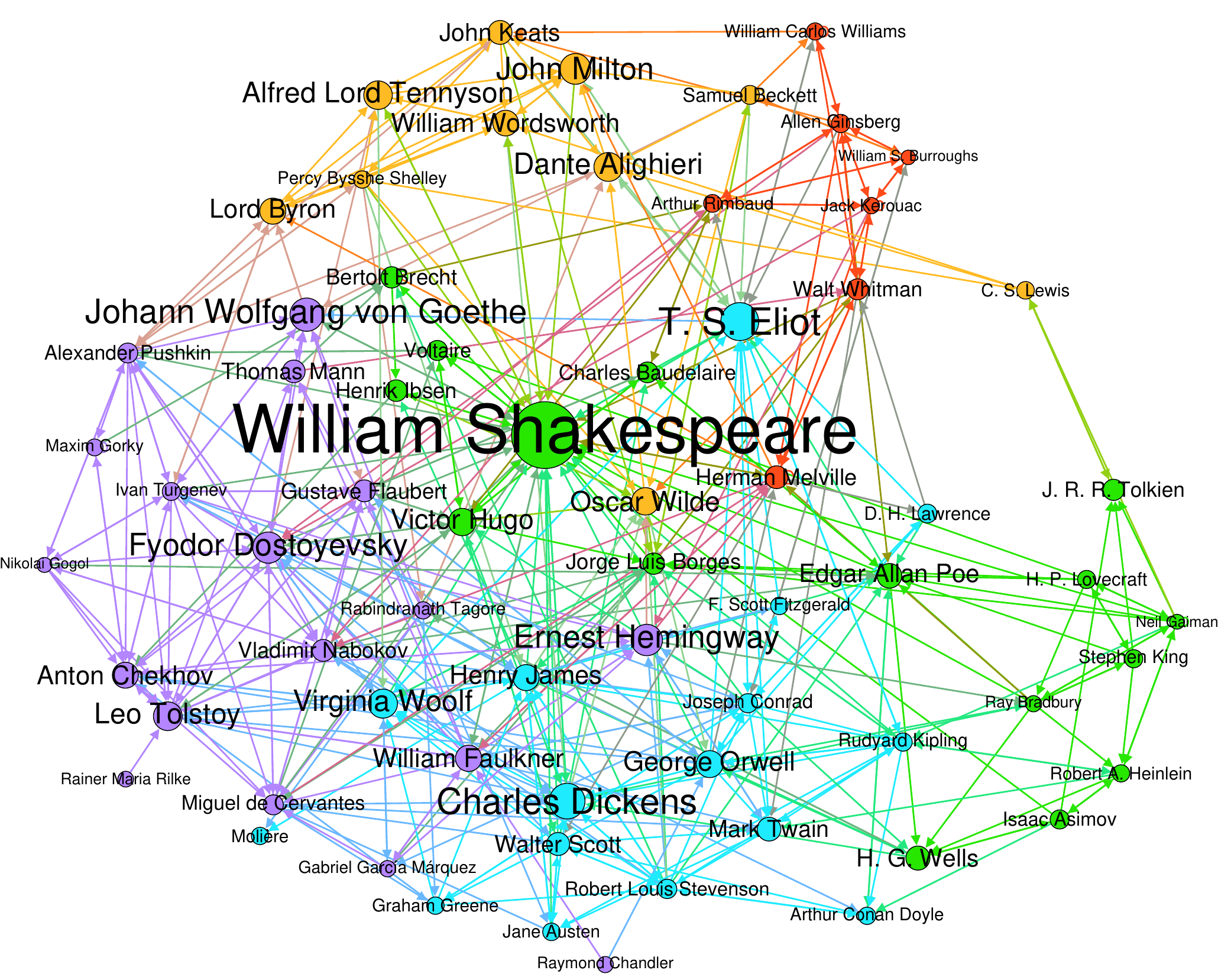}}
  \caption{Clustered link graph containing all writers with at least
60 in-links from other writers in our set.}\label{fig:influence}
\end{figure}

To visualize connections between writers, we created a graph of the
top writers in the English Wikipedia. The graph is depicted in
Fig.~\ref{fig:influence}. It was generated with Gephi using a
modularity-based clustering algorithm for coloring different
densely-connected subsets of writers. A writer $A$ is connected to a
writer $B$, if the article on $A$ contains a link to the article on
$B$.
The node and label size indicate the PageRank of the corresponding
writer by approach \rprwriter as
computed earlier in Section~\ref{sec:most-import-writ}.  For clarity,
we filtered nodes such that only writers with 60 or more incoming
links are shown.
%

Though the colored partition is not perfectly consistent, we notice
some interesting aspects. For instance, Russian writers like
\wpent{Dostoyevsky}, \wpent{Tolstoy}, or \wpent{Chekhov} are contained
in the purple part on the left side of the graph.
Similarly, three of the German writers, \wpent{Goethe}, \wpent{Thomas
  Mann}, and \wpent{Bertolt Brecht}, are also grouped together.
%
%
On the right side of the graph we find authors of fantasy novels, horror
and science fiction, like \wpent{J.\,R.\,R.~Tolkien},
\wpent{H.\,P.~Lovecraft}, \wpent{Stephen King}, \wpent{Ray Bradbury},
or \wpent{Isaac Asimov}. Their prominence on Wikipedia reveals a certain
bias among editors, and we can say the same thing about the fact
that only two women are contained in the graph, Jane Austen and Virginia
Woolf.  (To stress it once more, writers without a \wptype{Writer}
template at the time of creation of the DBpedia dump do not appear in
here, the likes of \emph{Homer}, \emph{Chaucer}, \emph{Proust},
\emph{Kafka}. To get them into the picture and create a more reliable
basic set of contributors to world literature requires future work on
the identification of writers.)

%

These few observations of interesting patterns already suggest that
this kind of graph can be used to group writers based on the
influences they exerted or received. Although the graph only contains
a small fraction of what we would call world literature, it has a
notable dimension and could be part of what \cite{levine2013what}
called a `timeless' and `deeply historical' world literature.
Nevertheless, a more comprehensive analysis is required, one that also
includes other language versions and considers other relationships
between writers, \eg the similarity between their categories.
Similar to \cite{athenikos2009philosophers}, we already tried to
include the \wptype{Writer} infobox properties \wpprop{influenced} and
\wpprop{influencedBy}, but due to their infrequent and inconsistent
use, the results were not meaningful. For example, Romanian poet
\wpent{Alexandru Macedonski} appeared to be exceedingly influential,
because in the used DBpedia dataset his infobox contains a huge list
of \wpprop{influenced} values, surely the work of an over-ambitious
editor.
%
%
Therefore, we decided not to use these two properties as they produce
misleading results in their current state.

\rja{is Taras Shevchenko a Russian writer? see the following URL to decide:
\url{https://en.wikipedia.org/wiki/Taras_Shevchenko\#Contribution_to_Russian_literature}}


\section{Transcending Language Boundaries}\label{sec:native-writers}

\rja{bei allen Captions/Column Headings der Abbildungen schauen, ob
  immer klar ist, was `language' meint (Sprache der Wikipedia-Version,
  oder Sprache des Autors?), z.B. in Table~\ref{tab:topfive}.}

%
Our last venture in this paper is to focus on the native languages
of writers (i.e., the languages they
have predominantly used in their works) and to compare them to the
languages of the Wikipedia editions in which they are ranked
prominently.
%
%
%
The assumptions underlying this analysis are:%
\begin{inparaenum}[(a)]
\item famous native-language writers should appear among the top
  positions in the rankings for ``their'' language edition,
\item writers whose works are considered world literature should also
  appear among the top positions of foreign-language editions
  (transgressional aspect).
\end{inparaenum}
Thereby, we again have the opportunity to compare the
proposed ranking approaches in terms of their suitability to
uncover the representation of world literature on Wikipedia.


In order to work with operable data, we manually identified the main
languages of our writers' works. More than one language was assigned
to writers who wrote major works in different languages (Nabokov, for
example, was assigned Russian and English).
For what it's worth, declaring \wpent{Virgil} and \wpent{Cicero}
`Italian' writers was probably the most daring intervention when
operationalizing our data. 
%
In the rankings in Table~\ref{tab:topfive} and on
\url{http://data.weltliteratur.net/} we have highlighted writers whose
writing language is equal to the corresponding Wikipedia language
edition.
The last column in Table~\ref{tab:topfive} indicates how many (and
which fraction) of the top 25 writers for each ranking can be
considered native-language writers of the corresponding language.
Since we have 7 rankings for each of the 15 language editions and
regard the top 25 writers for each such ranking, we can have at most
$7\cdot 15 \cdot 25 = 2,625$ distinct writers. But only 530 distinct
writers appear in all those rankings and only 29 of them have not
mainly written in one of the 15 languages. (Notably, these 530 are
more than the $15\cdot 25 = 375$ writers contained in one ranking,
thus the different approaches rank different writers to the top.)

The language of the writers will now be looked at from two directions:
first, by analyzing which writers are among the top positions in the
Wikipedia edition of ``their'' language, and second, by analyzing which
writers transcend language boundaries and have top ranks in other
language editions.
We assess both aspects quantitatively and qualitatively.


\subsection{Writers in Their Own Language}\label{sec:writers-their-own}

\enlargethispage{\baselineskip}

We start with a quantitative analysis by counting how many
native-language writers are among the top 25 writers for each ranking
measure and language edition. Table~\ref{tab:approaches_comp} shows for
each ranking measure the mean number of native-language writers over
all language editions.
%
%
%
%
%
\begin{table}
  \centering
  \caption{\bf Mean number of native-language writers.}\label{tab:approaches_comp}
  \begin{tabular}{@{}lr@{$\pm$}r@{ }l@{}}
    \toprule
    measure                   &  \multicolumn{3}{l@{}}{native-language writers} \\
    \midrule
    page length               &8.87  &4.18 &(35.5\%) \\
    number of in-links        &10.07 &3.87 &(40.3\%) \\
    PageRank writers          &4.87  &5.24 &(19.5\%) \\
    PageRank complete         &9.80  &4.83 &(39.2\%) \\
    number of page views 2012 &14.33 &4.38 &(57.3\%) \\
    number of page views 2013 &14.20 &4.46 &(56.8\%) \\
    number of page views 2014 &14.53 &4.15 &(58.1\%) \\  
    \bottomrule
  \end{tabular}
  \begin{flushleft}
    For each approach we list the mean number of native-language writers
    (including standard deviation and percentage) among the top 25
    writers.
  \end{flushleft}
\end{table}


The high fraction of native-language writers among the top 25 writers in the
ranking by \emph{page views} (around 57\%) shows that native-language
writers obviously are very important for readers of the corresponding
language editions.
Again, the \emph{number of in-links} and the \emph{PageRank writers}
exhibit similar results with around 40\% native-language writers.
This is also the highest value among the intrinsic ranking measures,
closely followed by the \emph{page length}. This shows that
native-language writers are well-represented on Wikipedia, i.e.,
they have comprehensive articles and they are mentioned in other
articles.

%
Considering Table~\ref{tab:topfive} again, a surprising result is that
only Romanian writers are among the top five in the ranking by
\emph{page length} for the English Wikipedia. This shows in all
clarity how much this ranking depends on diligent editors. It is
certainly not usable to identify important writers.
Surprisingly long articles can be found on other Wikipedia editions as
well, \eg \wpent{Agatha Christie} ranks 3rd in Portuguese and
\wpent{George Orwell} 2nd in Arabic.
All in all, 133 out of 375 writers (the top 25 for the 15 languages)
match ``their'' language, although they often cannot be considered the
most prominent writers for the corresponding
language. 
(Cf. \url{http://data.weltliteratur.net/ranking.html\#page-length}
  for details.) 
The ranking based on the \emph{number of in-links} places a
native-language writer on the first rank in 7 out of 15 languages
while this is the case for only 2 writers in the \emph{PageRank
  writers} ranking. If we leave out Shakespeare, 12 native-language
writers are on the first rank regarding the \emph{number of in-links}
and 7 regarding the \emph{PageRank writers}.  Among the top five, more
than half (40) of the 75 writers match ``their'' language when ranked
by the \emph{number of in-links} while this is true for around a
quarter (20) of the writers in the \emph{PageRank writers} ranking.
Although \wpent{Shakespeare}, \wpent{Dickens} and \wpent{Tolkien} are
certainly important English-language writers, \wpent{Robert Christgau}
and \wpent{Roger Ebert} do not really fit into the list.
Since both are critics, their articles receive many incoming
links from articles on the many works they have reviewed --
\wpent{Ebert} from articles on movies, \wpent{Christgau} from
articles on music albums 
(e.g., of the 5,125 articles that
  link to \wpent{Robert Christgau}, 1,705 have the string ``album'' in
  their title and also most of the other ones seem to refer to albums
  or songs, too.). 
%
%
%
But, altogether, we are able to identify important authors by counting
the \emph{number of in-links} and are also able to rank, in many
cases, important writers to the top positions for the corresponding
language.
As we have seen in Section~\ref{sec:ranking-correlation}, the ranking
by \emph{PageRank complete} is numerically similar to the ranking by
the \emph{number of in-links}, but there is an apparent difference:
the two aforementioned critics are no longer among the top five,
instead, \wpent{Goethe} and \wpent{Cicero} enter the picture.

\enlargethispage{\baselineskip}
Although we expected the ranking by \emph{PageRank writers} to
outperform the \emph{number of in-links} approach, this is mostly not
the case: the \emph{PageRank writers} can identify popular writers,
but they typically do not fit to the corresponding language. It
therefore is better suited to identify writers that transcend
languages -- an aspect we investigate in
Section~\ref{sec:writ-other-lang}.
One possible explanation could be that articles of native-language
writers contain more references and thereby more outgoing links which
in return reduces their PageRank.


%
%





The rankings by the \emph{number of page views} also have important
native-language writers among their top positions, but we can also
observe that they reflect current events. For instance, German writer
and publisher \emph{Frank Schirrmacher} (who died in 2014) is among
the top 10 in the German Wikipedia in 2014 and so is Polish-German
literary critic and writer \emph{Marcel Reich-Ranicki} in 2013 (also
the year he died).
Nevertheless, there are big constants, for example the three German
writers \emph{Goethe}, \emph{Schiller},
and \emph{Brecht} who are among the top five in all three
rankings of the German edition.
\rja{Frank: are there/do we need good examples from other languages?}

\subsection{Writers in Other Languages}\label{sec:writ-other-lang}


Let us now change our perspective and consider the most important
writers in Wikipedia language editions different from their writing
languages, i.e., writers who transcend language boundaries. This is an
interesting new perspective and adds to other methods for measuring
the cross-lingual impact of authors, like counting the number of
translations.

For each writer, we compute the sum of the reciprocal ranks over all
language editions except the native language of the writer. E.g.,
German writer \emph{Bertolt Brecht} has a score of $0.94 = 1/19 + 1/9
+ 1/19 + 1/20 + 1/13 + 1/10 + 1/2$ for the \emph{number of in-links}
ranking which results from a 19th, 9th, 19th, 20th, 13th, 10th, and
2nd place in the English, French, Russian, Spanish, Portuguese,
Hungarian, and Turkish Wikipedia, respectively.
We then rank all writers with the same native language according to
this score and repeat this for the other ranking measures.  This
results in one ranking for each ranking measure and language.
\begin{table}
    \caption{\bf Writers ranking highest in foreign-language editions.}
  \label{tab:top5-native}
  {\footnotesize 
  \setlength{\tabcolsep}{2pt}
  \begin{tabular}{@{}llll@{}}
    \toprule


    rank                    &English                        &German                            &French             \\
    \midrule
    1st                     &William Shakespeare (10.33)    &Johann Wolfgang von Goethe (2.40) &Voltaire (1.18)    \\  
    2nd                     &J.\,R.\,R. Tolkien (1.94)      &Friedrich Schiller (0.26)         &Victor Hugo (0.84) \\   
    3rd                     &Edgar Allan Poe (0.89)         &Bertolt Brecht (0.12)             &Molière (0.82)     \\
    4th                     &Mark Twain (0.64)              &Thomas Mann (0.06)                &Jules Verne (0.06) \\
    5th                     &Charles Dickens (0.61)         &Karl May (0.04)                   &Émile Zola (0.05)  \\
    count\textsubscript{25} &25                             &5                                 &6                \\
    \midrule
    rank                    &Russian                        &Italian (Roman)                   &Spanish                     \\
    \midrule                
    1st                     &Leo Tolstoy (0.68)             &Cicero (3.34)                     &Isabel Allende (1)          \\
    2nd                     &Fyodor Dostoyevsky (0.60)      &Dante Alighieri (1.25)            &Miguel de Cervantes (0.18)  \\
    3rd                     &Anton Chekhov (0.28)           &Virgil (0.86)                     &Reinaldo Arenas (0.09)      \\
    4th                     &Alexander Pushkin (0.09)       &Ovid (0.38)                       &--                           \\
    5th                     &Constantin Stanislavski (0.06) &Petrarch (0.11)                   &--                           \\
    count\textsubscript{25} &5                              &6                                 &3                           \\

    \bottomrule
  \end{tabular}
  } 
  \begin{flushleft}
    Top five native writers for English, German, French, Russian,
    Italian, and Spanish ranking highest in the other 14
    foreign-language editions, respectively, according to
    \emph{PageRank complete}. The last row (count\textsubscript{25})
    indicates how many native writers of each language are among the
    top 25 writers in \emph{PageRank complete} rankings of the remaining
    14 language editions.
  \end{flushleft}
\end{table}
Table~\ref{tab:top5-native} shows the top five native-language
writers for six languages that appear most prominent
among the top~25 writers in the rankings of the
other 14 languages according to \emph{PageRank
  complete}. 
%
For the remaining nine languages none of their native writers appear
among the top~25 in other languages in those rankings.


The results show which foreign-language writers are well-represented
in terms of links to their articles in the different editions of Wikipedia.
There is a clear bias towards Western
culture which likely is also induced by the selection of the 15
languages (of which 9 can be regarded to represent Western culture,
give or take).
Most of the writers in Table~\ref{tab:top5-native} also appear
among the top~25 in the rankings by \emph{PageRank complete} of
their native language, except for \emph{Isabel Allende},
and \emph{Reinaldo Arenas}.
\emph{Allende} ranks 1st in the Serbo-Croatian and
\emph{Arenas} 11th in the Persian Wikipedia, so they
might bear a special importance in those language editions.

The results for each of the other six rankings can be found at
\url{http://data.weltliteratur.net/ranking_native.html}.
%
Comparing the results of the intrinsic \emph{PageRank complete}
measure shown in Table~\ref{tab:top5-native} to the extrinsic
\emph{number of page views} measure on the web page, we can observe
differences and commonalities:
\begin{itemize}
\item Deaf-blind American writer and activist \emph{Helen Keller} is
  ranked much higher by the number of page views (always among the
  top~five) than by the \emph{PageRank complete} (not among the
  top~25). So her article is consulted by many people,
  but she is not that well integrated in the network of
  writers on Wikipedia.
\item According to \emph{PageRank complete}, \emph{Johann Wolfgang von
    Goethe} is the top German writer in other language editions,
  followed by \emph{Schiller}, \emph{Brecht}, \emph{Thomas Mann}, and
  \emph{Karl May}.  Page-view numbers from 2012 to 2014 also see
  \emph{Goethe} upfront, but this time followed by \emph{Hermann
    Hesse} and \emph{Erich Maria Remarque}, showing different
  priorities of editors and readers.
\item French writers \emph{Victor Hugo}, \emph{Jules Verne}, and
  \emph{Voltaire} constantly rank 1st, 2nd, and 3rd over the
  three focal years.
\item The first four Russian writers are also always the first four by
  the page-view measure.
\item Roman/Italian writers \emph{Cicero} and \emph{Dante Alighieri}
  are also always the first two.
\item The only Spanish-language writer among the top~25 by page views is
  \emph{Gabriel García Márquez} who is not among the top~25 by
  \emph{PageRank complete}.
\end{itemize}
\rja{Frank: wollen wir das so als Liste lassen? Müssen wir die
  Ergebnisse noch mehr diskutieren/interpretieren?}
These are just some few observations, but they already show how
conveniently we can tell the difference between what editors and what
readers find important. The project website features much more
incentives to take a closer look at the representation of world
literature on Wikipedia.


\section{Summary and Conclusion}\label{sec:conclusion}

\rja{deutlich erweitern. Themen z.B.:
- Wikipedia Demographics / Bias (``junge weiße Männer'')
- andererseits: Nachteil der bestehenden (Verkaufszahl/Kritiker-basierten) Methoden, Vorteil Wikipedia: breite Masse
- Wozu das ganze?
- Was ist der Impact (für DH/für CS)?
- Wo sind die Grenzen?
}




In this method-oriented study, we presented a framework that can unlock
Wikipedia as a new source for research on world literature. The methods
we suggested and discussed are also meant to contribute to the growing
tool set of the Digital Humanities and, more precisely, the emerging
field of Digital Literary Studies.

Wikipedia itself is subject to constant change and so are the formalized
datasets derived from it.
In any case, the current limitations we discussed leave room for
improvement.
Our results show that although Wikipedia provides different options to
classify articles on writers, the identification of writers of literature
across different Wikipedia language versions is a challenging task,
partially due to the inconsistent use of the \wptype{Writer} template.
%
%
However, we were able to create sets that 
comprise the majority of famous writers of literature contained in the
English Wikipedia, while presumably containing just few non-writers.
%
%
%

The most eminent author in all languages we considered is by far
\wpent{William Shakespeare}. We find him in the top five of every
analyzed language version for three of the chosen measures, even placed
first in 9 of 15 languages based on the number of in-links and in 13
of 15 versions based on the PageRank between writers of our
set. Shakespeare's central position in world literature, his strong
influence on many writers around the world is not news, of course, but
we confirmed this with an unconventional method, a method that
produced many more insightful results.

There are three main things we can feed into the ongoing discourse on
world literature. First, a specific Wikipedia language version tends
to lay emphasis on the most eminent literary eras of the respective
language. The preoccupation of a Wikipedia language version with its
own classical authors can be interpreted as a possible precondition
for these authors to eventually become a part of world literature.
Second, our analysis of the
top 25 writers in 15 language versions shows which writers actually
crossed national and language borders and have retained a significant
presence in foreign nations. The strength of this approach is that it
both confirms the more or less intuitive knowledge of which writers
are most influential and provides a much-needed nuance to the
specific constellations of world literature in particular languages.
Third, the network graph of the top writers
in the English Wikipedia offers a new approach to present how world
literature is currently represented in Wikipedia.
%
%
%
%



%

In this paper, our analysis is restricted to writers. The logical next
step would be to develop an approach to reliably identify
\emph{literary works} of note. Although the methods presented in
Section~\ref{sec:approaches} could be applied by identifying
appropriate Wikipedia templates and categories for literary works, a
preliminary analysis showed that there are other alternatives. E.g.,
in some cases there are dedicated pages for some writers that list
their works, as demonstrates \cite{wp:shakespearbibliography}, 
or the works are listed on the writer's page themselves, e.g., \cite{wp:irving}. 
%
Furthermore, a more fine-grained and coherent classification of
writers into fiction, non-fiction writers, etc. within Wikipedia could
help to answer the question \emph{Who is a writer (of literature)?}
That way, we could divide our basic set into sets of different types
of writers.
The analysis of the editing history of articles could also proof relevant
for a definition of world literature in order to gain insights into
actual editorial work that leads to the current representation of world
literature within this growing repository of human knowledge production.
An in-depth comparison and qualitative analysis of the different
prominence measures could help to figure out whether the measures rank
different types of writers to the top.

Wikipedia is still modified every few seconds, DBpedia sets are
released twice a year. It is our plan to build an observatory
around the measures introduced in this paper so that we can observe this
ever-changing entity called world literature through the eye of
the tens of thousands of Wikipedia editors and millions of readers.

Last not least, studies like ours could advocate for scholars to
actively improve the quality of articles, add missing pieces and
help repair structural shortcomings, especially with regard to the rather
conservative version of world literature that Wikipedia implicitly
conveys for the time being.

\section*{Acknowledgements}


We would like to thank Constanze Baum, Florian Lemmerich and Michel De
Dobbeleer for their valuable feedback.


\bibliography{bibliography}


\end{document}